\documentclass[useAMS,usenatbib]{mnras}
\usepackage{graphicx}
\usepackage{amsmath}
\usepackage{amssymb}
\usepackage{color}
\usepackage{mathptmx}
\usepackage{bm}
\definecolor{caribbeangreen}{rgb}{0.0, 0.8, 0.6}
\definecolor{darkpastelgreen}{rgb}{0.01, 0.75, 0.24}
\definecolor{guppiegreen}{rgb}{0.0, 1.0, 0.5}
\definecolor{palatinateblue}{rgb}{0.15, 0.23, 0.89}
\definecolor{violet}{rgb}{0.56, 0.0, 1.0}

\title[Local photoionization feedback effects on galaxies]{Local photoionization feedback effects on galaxies}
 \author[Obreja et al.]{Aura Obreja$^{1,2}$\thanks{E-mail: obreja@usm.lmu.de}, Andrea V. Macci\`{o}$^{2,3}$, Benjamin Moster$^{1,4}$, Silviu M. Udrescu$^{2}$, \and Tobias Buck$^{5}$, Rahul Kannan$^{6}$\thanks{Einstein Fellow}, Aaron A. Dutton$^{2}$ and Marvin Blank$^{2,7}$\\ 
 $^{1}$University Observatory Munich, Scheinerstra\ss e 1, D-81679 Munich, Germany\\
 $^{2}$New York University Abu Dhabi, PO Box 129188, Saadiyat Island, Abu Dhabi, United Arab Emirates\\ 
 $^{3}$Max-Planck-Institut f\"{u}r Astronomie, K\"{o}nigstuhl 17, 69117 Heidelberg, Germany\\
 $^{4}$Max-Planck-Institut f\"{u}r Astrophysik, Karl-Schwarzschild Stra\ss e 1, 85748 Garching, Germany\\
 $^{5}$Leibniz-Institute for Astrophysics Potsdam, An der Sternwarte 16, 14482 Potsdam, Germany\\
 $^{6}$Harvard-Smithsonian Center for Astrophysics, 60 Garden Street, Cambridge 02138, Massachusetts, USA\\
 $^{7}$Institut f\"{u}r Theoretische Physik und Astrophysik, Christian-Albrechts-Universit\"{a}t zu Kiel, Leibnizstr. 15, D-24118 Kiel, Germany}
 \begin{document}

\maketitle

\label{firstpage}

\begin{abstract}
We implement an optically thin approximation for the effects of the local radiation field from stars and hot gas on the gas heating and cooling in the N-body SPH code {\sc gasoline2}. We resimulate three galaxies from the NIHAO project: one dwarf, one Milky Way-like and one massive spiral, and study what are the local radiation field effects on various galaxy properties. We also study the effects of varying the Ultra Violet Background (UVB) model, by running the same galaxies with two different UVBs. Galaxy properties at $z=0$ like stellar mass, stellar effective mass radius, HI mass, and radial extent of the HI disc, show significant changes between the models with and without the local radiation field, and smaller differences between the two UVB models. The intrinsic effect of the local radiation field through cosmic time is to increase the equilibrium temperature at the interface between the galaxies and their circumgalactic media (CGM), moving this boundary inwards, while leaving relatively unchanged the gas inflow rate. Consequently, the temperature of the inflow increases when considering the local radiation sources. This temperature increase is a function of total galaxy mass, with a median CGM temperature difference of one order of magnitude for the massive spiral. The local radiation field suppresses the stellar mass growth by ∼20 per cent by $z=0$ for all three galaxies, while the HI mass is roughly halfed. The differences in the gas phase diagrams, significantly impact the HI column densities, shifting their peaks in the distributions towards lower $N_{\rm HI}$.
\end{abstract}


\section{Introduction}
\label{intro} 

Galaxy formation is tightly linked with the reionization history. While gravity and gas cooling processes 
cause dark matter haloes and galaxies to form, their stars and quasars (QSOs) are the sources of primary 
ionization photons that heat up and ionize their surrounding medium, thus providing a negative feedback for further 
gas inflow, especially in the dwarf galaxy regime \citep{Efstathiou:1992}. Given the large dynamical range of all 
the relevant processes, cosmological simulations provide one of the most powerful methods to theoretically study 
galaxy formation and reionization. However, solving the radiation transfer equation simultaneously with evolving 
structure formation, hydrodynamical equations and star and black hole formation plus their feedback on the gas 
still has a prohibitive computational cost 
\citep[][]{Altay:2008,Pawlik:2008,Petkova:2009,Petkova:2011,Rosdahl:2013,Ocvirk:2016,Kannan:2019}. 
For this reason, theoretical studies tend to focus either on reionization while treating galaxy formation in a very approximate manner, 
or on galaxy formation assuming an isotropic and time dependent radiation field, the so-called Ultra Violet 
Background \citep[UVB, e.g.][]{Haardt:1996,Faucher:2009,Haardt:2012} used in cosmological simulations.

The UVB is set-up by all the high energy photons (E$>$13.6~eV) emitted 
from primary (stars and QSOs) and secondary \citep[highly ionized gas surrounding stars and QSOs,][]{Haardt:1996} sources, that can escape their 
host dark matter haloes, throughout the history of the universe. The UVB determines the thermal and ionization state of diffuse 
intergalactic medium (IGM) and the build-up of the various UVB sources also directly determines the HI and He II reionizations. 
The HI reionization was completed by $z\sim5$ \citep[e.g.][]{Cen:1993,Fukugita:1994,Schenker:2014,Becker:2015}, 
and it is thought to have been mainly driven by the UV radiation from massive stars \citep[e.g.][]{Shapiro:1987,Madau:1999}, 
though a high-$z$ faint QSO population could have also had a major contribution to it \citep[e.g.][]{Miralda-Escude:1990,Madau:2015}. 
The HeII reionization was completed later on at $z\sim2.8$ \citep[e.g.][]{Shull:2004,Worseck:2011,Becker:2011,Hiss:2018}, 
and it is thought to have been driven by the hard UV radiation from QSOs \citep[e.g.][]{Kriss:2001,Compostella:2013}.

The UVB models are an essential part of galaxy formation studied through cosmological simulations. Their z-dependent tabulated 
values of hydrogen and helium photoionization and photoheating rates are a direct input for the equations governing the cooling and heating of the gas. 
The model of \citet{Haardt:1996} was among the first UVB model to be incorporated into cosmological simulations 
\citep[e.g.][]{Weinberg:1997,Kallander:1998,Borgani:2004}. Over the years, these authors have updated their original model 
\citep[][hereafter HM12]{Haardt:2012}, and few other groups have proposed alternatives 
\citep[e.g.][]{Faucher:2009,Faucher:2019,Onorbe:2017,Khaire:2019,Puchwein:2019}. 
Virtually all recent cosmological simulations of galaxy formation still use either versions of the Haardt \& Madau UVB 
\citep[e.g.][]{Wang:2015,Schaye:2015} or \citet[][hereafter FG09]{Faucher:2009} \citep[e.g.][]{Grand:2017,Hopkins:2018b}.

The very recent UVB model of \citet{Khaire:2019} is being added into the newest version (17.01) of the CLOUDY spectral synthesis 
code \citep{Ferland:2013}, and is based on up-to-date observational data on emissivities of type-1 QSO and galaxies, HI distribution
in the IGM, and average escape fractions of ionizing HI photons from galaxies. \citet{Khaire:2019} also show how the two most used UVB 
models in simulations, HM12 and FG09, compare with one another, and with the very recent data used them to constrain their own model. 
E.g., in terms of photoionization rates, FG09 results in lower HI rates than HM12 between $z\sim0.5$ and $z\sim5.8$, and 
significantly lower HeII rates below $z\sim3.2$. The HI and HeII rates of FG09 are higher than the ones of HM12 at 
redshifts $\gtrsim 6$ and $\gtrsim 3.5$ respectively, 
where there are few observational constraints \citep[e.g.][]{Yue:2016}. 
The most recent UVB models \citep[e.g.][]{Khaire:2019,Puchwein:2019} conclude that HM12 underestimate the 
$z=0$ HI photoionization rate by a factor of $\rm\sim2$.
At lower redshifts ($z<6$) where tighter observational constraints exist, the reconstructed UVB Spectral Energy Distribution (SED) 
in the HM12 model has a larger amplitude than the one of FG09 at most relevant wavelengths.   
The assumed UVB models have a particularly high impact on the formation of less massive galaxies, a strong UVB being able to suppress 
SF in these low mass systems \citep[e.g.][]{Efstathiou:1992,Weinberg:1997}. Also, due to the very poor observational constraints for 
$z>6$, the UVB models can vary significantly between each other at high redshifts, and therefore change the predictions for the 
low end of the high-$z$ luminosity function \citep[e.g.][]{Samui:2007}.

Reionization is actually a non spatially homogeneous process and on galaxy scales the high energy photon distribution 
can be dominated by the anisotropic local radiation sources like stars, black holes and hot gas. A patchy reionization and 
a strong inhomogeneous galactic radiation field are bound to affect the galaxy formation process.
\citet{Kannan:2014} were the first to study the effect of the local radiation sources on the formation of a Milky Way mass galaxy 
in a cosmological simulation. These authors got inspired by the analytical models of \citet{Cantalupo:2010} and \citet{Gnedin:2012}
to develop an optically thin approximation to account for the effects of ionizing radiation from SF regions and 
post-asymptotic giant branch stars (AGB) on the thermal and ionizing state of the gas.  
Kannan et al's radiation field implementation in the old version of the N-body smoothed particle hydrodynamics (SPH) code {\sc gasoline} 
\citep{Wadsley:2004} is an approximation that does not increase significantly the computational time such that zoom-in simulations of galaxies 
can be evolved to $z=0$. This is not the case with hydrodynamical codes that actually solve the radiation transfer equation on the fly 
\citep[e.g.][]{Pawlik:2008,Kannan:2019}. Optically thin approximation means that: the radiation field at each 
position in the simulation domain scales as the inverse square distance to the source, and all absorption effects are 
quantified through source dependent escape fractions. 

\citet{Cantalupo:2010} used the fact that star formation rate (SFR) correlates 
obsevationally with the soft X-ray flux \citep[e.g.][]{RosaGonzalez:2009} to show how the gas cooling curve and implicitly 
cooling time changes as a function of SFR for a range of virial halo masses. \citet{Cantalupo:2010} also showed how these SFR dependent 
cooling curves differ from the one computed when the only photoionization source considered is the UVB. Given that soft X-rays are 
capable of ionizing various metal species, a large enough flux can significantly increase the gas cooling time, 
therefore acting like a preemptive feedback mechanism; gas is not ejected from the galaxy like in the explosive 
feedback associated with supernovae (SN) events \citep[e.g.][]{Stinson:2006,DallaVecchia:2008}, but is prevented from cooling and hence from 
forming stars. \citet{Gnedin:2012} proposed to describe the radiation field from stars and active galactic nuclei (AGN) on galaxy scales through 
a general functional form based only on a few parameters. These authors also showed that the effects of such ionization fields on the heating 
and cooling functions can be described as functions of only a few important metal species. 
The original radiation field implementation of \citet{Kannan:2014}
was further refined by \citet{Kannan:2016} for the {\sc arepo} code \citep{Springel:2010} 
to include the Bremsstrahlung from hot halo gas, and tested on prepared simulation of a few galaxy clusters.  

\citet{Hopkins:2018a} use two schemes to study radiation feedback effects in galaxies from ultra 
faint dwarfs to Milky Way-likes: a ray-based and a moments-based method. The ray-based method assumes 
all absorbtion effects to occur in the vicinity of the radiation sources and negligible light travel times, and 
therefore it represents an optically thin approximation to radiation transport, same as the implementation of 
\citet{Kannan:2014,Kannan:2016}. The moments-based method of \citet{Hopkins:2018a} is intended 
to capture the optically thick regime and is photon-conserving by construction. However, it requires 
a strict enforcement of the finite light speed, increasing the computational cost, and does not converge to 
the correct solution for the radiation transfer equation in the optically thin case.

Currently, there are thus only two works, \citet{Kannan:2014} and \citet{Hopkins:2018a}, that studied the 
effects of the local 
radiation field on cosmological simulations of galaxy formation and evolution, on a combined sample of seven galaxies 
ranging from dwarfs to L$^{\rm *}$. \citet{Kannan:2014} found a significant decrease in stellar mass and peak
circular velocity at $z=0$ for one Milky Way like galaxy when the local radiation field was taken into account.
\citet{Hopkins:2018a} reach the opposite conclusion from their two simulated L$^{\rm *}$ galaxies: 
no change in stellar mass or peak circular velocity, but found more significant effects for their four simulated 
dwarf galaxies. Given the paucity of such studies, here we explore the effects of the local photoionization 
and photoheating produced by stars and hot gas and the effects of varying the UVB model on three galaxies initially 
simulated within the project Numerical Investigation of a Hundred Astrophysical Objects \citep[NIHAO,][]{Wang:2015}.  
For this purpose we have implemented the optically thin approximation of \citet{Kannan:2014,Kannan:2016} in the N-body SPH 
code {\sc gasoline2} \citep{Wadsley:2017}. 

This paper is organized as follows. Section~\ref{heat_cool} describes how gas heating and cooling are treated in the code, while 
Section~\ref{section_lpf} presents the implementation of the local radiation field sources in the gravity structure of {\sc gasoline2}.
Section~\ref{sims} gives a brief overview of the three NIHAO galaxies resimulated with the alternative FG09 UVB with and without 
the extra contribution from the local radiation field. The results are presented in Section~\ref{results}, subdivided into: a 
discussion on the differences in formation histories, and one on the $z=0$ galaxy properties. The summary and discussion of results 
are given in Section~\ref{conclusions}.

\section{Heating and cooling in {\sc gasoline2}}
\label{heat_cool}

Most cosmological simulations of galaxy formation include radiation effects as a redshift-dependent homogeneous UVB which acts as 
photoionization and heating source in the hydrodynamic energy equation. Explicitly, this means that the heating ($H$) and cooling 
($\Lambda$) of each gas particle/cell $k$ will be functions of density ($\rho$), temperature ($T$), composition globally described by total 
metallicity ($Z$), and radiation field ($J$): 

\begin{align}
\label{heat_cool_eq}
H_k &= H_k(\rho_k,T_k,Z_k,J)\\ \nonumber
\Lambda_k &= \Lambda_k(\rho_k,T_k,Z_k,J) \nonumber
\end{align}

The thermal and ionization state of the gas is set by the balance between heating and cooling. In practical terms, for each gas element 
(cell or particle) the compact forms for $H$ and $\Lambda$ actually imply solving a large set of coupled differential equations for the time 
evolution of the number density $n_{\rm(i)}$ of all atomic and molecular species present (i). Given the high complexity of these so-called 
chemical networks, most cosmological simulations codes, including {\sc gasoline2} \citep{Wadsley:2017} as described in \citet{Shen:2010}, 
solve on the fly only the non-equilibrium equations governing the ionization states of hydrogen and helium (HI, HII, HeI, HeII, HeIII, 
e$^{\rm-}$):
\begin{align}
\label{eq_ionization}
 \frac{dn_{\rm HI}}{dt} =& \alpha_{\rm HII}n_{\rm HII}n_e - \Gamma_{\rm coll \, HI}n_{\rm HI}n_e - \Gamma_{\rm phot \, HI}n_{\rm HI}\\ \nonumber
 \frac{dn_{\rm HeI}}{dt} =& (\alpha_{\rm HeII}+\alpha_{\rm diel})n_{\rm HeII}n_e \\ \nonumber
                    & -\Gamma_{\rm coll \, HeI}n_{\rm HeI}n_e - \Gamma_{\rm phot \, HeI}n_{\rm HeI}\\ \nonumber
 \frac{dn_{\rm HeII}}{dt} =& - \frac{dn_{\rm HeI}}{dt} + \alpha_{\rm HeIII}n_{HeIII}n_e \\ \nonumber
                        & - \Gamma_{\rm coll \, HeII}n_{\rm HeII}n_e - \Gamma_{\rm phot \, HeII}n_{\rm HeII}\\ \nonumber
\end{align}
where $\Gamma_{\rm coll \, (i)}$ and $\Gamma_{\rm phot \, (i)}$ are the collisional ionization and photoionization rates of ion species (i),
$\alpha_{\rm(i)}$ are the radiative recombination coefficients, and $\alpha_{\rm diel}$ is the dielectric recombination coefficient.
In {\sc gasoline2}, we use the collisional ionization rates from \citet{Janev:1987}, the radiative recombination rates from \citet{Verner:1996}, 
and the dielectric recombination coefficient from \citet{Aldrovandi:1973}. 

To obtain a set of closed equations for the evolution of the ion species, we use the relation between helium abundance and total 
metallicity of \citet{Jimenez:2003} and the following two conservation equations:
\begin{align}
 \label{eq_cons}
 n_{\rm H} =& n_{\rm HI} + n_{\rm HII} \\ \nonumber
 n_{\rm e} =& n_{\rm HII} + n_{\rm HeII} + 2n_{\rm HeIII}\\ \nonumber
\end{align}
In practice, we compute the photoionization rates for hydrogen and helium ion species as:
\begin{align}
\label{phot_rates}
 \Gamma_{\rm phot \, (i)} =& \int_{\rm\nu_{\rm lim(i)}}^{\rm \infty}\frac{4\pi J_{\rm\nu}^{\rm(UVB)}}{h\nu}\sigma_{\rm\nu_{\rm (i)}}d\nu + \int_{\rm\nu_{\rm lim(i)}}^{\rm \infty}\frac{4\pi J_{\rm\nu}^{\rm(LPF)}}{h\nu}\sigma_{\rm\nu_{\rm (i)}}d\nu\\ \nonumber
 =& \Gamma_{\rm phot \, (i)}^{\rm(UVB)} + \Gamma_{\rm phot \, (i)}^{\rm(LPF)}
\end{align}
where $\nu_{\rm lim(i)}$ is the ionization frequency for species (i), $\sigma_{\rm\nu_{\rm (i)}}$ is the frequency dependent 
photoionization cross-section, $h$ is the Planck constant, and the incident radiation field $J_{\rm\nu}$ is separated into a contribution 
from the UVB and a contribution from the local photoionization field. The photoionization rates $\Gamma_{\rm phot \, (i)}^{\rm(UVB)}$ in 
Equation~\ref{phot_rates} are usually interpolated from redshift-dependent homogenous UVB models.  
In this work we use both HM12 and FG09 UVB models, to study how the choice of a particular model influences the observable 
properties of simulated galaxies. We will come back to how exactly we compute the photoionization rates contribution 
$\Gamma_{\rm phot \, (i)}^{\rm(LPF)}$ in the following section.

The photoheating rate $H$ that is needed in the hydrodynamic energy equation is computed as:
\begin{equation}
\label{heating_eq}
 H = n_{\rm HI}\epsilon_{\rm HI} + n_{\rm HeI}\epsilon_{\rm HeI} + n_{\rm HeII}\epsilon_{\rm HeII} + H_{\rm metal}
\end{equation}
where $H_{\rm metal}$ is the photoheating rate due to elements heavier than helium, dependent on the full radiation field $J=J^{\rm(UVB)}+J^{\rm(LPF)}$.  
The photoheating rate density $\epsilon_{\rm(i)}$ of each primordial ion species (i) is calculated as:
\begin{align}
\label{photheat_eq}
 \epsilon_{\rm(i)} =& \int_{\rm\nu_{\rm lim(i)}}^{\rm \infty}\frac{4\pi J_{\rm\nu}^{\rm(UVB)}}{h\nu}\sigma_{\rm\nu_{\rm (i)}}(h\nu - h\nu_{\rm lim(i)})d\nu \\ \nonumber
 & +\int_{\rm\nu_{\rm lim(i)}}^{\rm \infty}\frac{4\pi J_{\rm\nu}^{\rm(LPF)}}{h\nu}\sigma_{\rm\nu_{\rm (i)}}(h\nu - h\nu_{\rm lim(i)})d\nu\\ \nonumber
 =& \epsilon_{\rm(i)}^{\rm(UVB)} + \epsilon_{\rm(i)}^{\rm(LPF)}\nonumber
\end{align}
As in the case of the photoionization rate, the photoheating rate density of the primordial ion species is 
separated into a term due to the UVB, $\epsilon_{\rm(i)}^{\rm(UVB)}$ being interpolated from tables provided 
by the UVB models, and one due to the local radiation field, $\epsilon_{\rm(i)}^{\rm(LPF)}$. 
In the current version of {\sc gasoline2}, the SPH energy equation has an additional 
\emph{heating} term, not included in $H$ of Equation~\ref{heating_eq}, 
which mimicks the energy input from massive stars prior 
to their SNe II phase into the surrounding gas, effect also known as `early stellar feedback' 
\citep{Stinson:2013a}. This energy from massive stars close to the end of their lifetimes is distributed over the 
SPH kernel, and therefore it is a short range effect \citep[see also][]{Rosdahl:2015,Kannan:2018}. Complementary, 
$\epsilon_{\rm(i)}^{\rm(LPF)}$ models the (photo)heating on large scales.

Following \citet{Shen:2010}, the cooling rate $\Lambda$ is a linear combination of three terms: one due to hydrogen and helium line cooling, 
$\Lambda_{\rm H,He}$ which is solved for on the fly assuming non-equilibrium, one due to Compton scattering of the Cosmic Microwave Background (CMB) 
photons, and the last due to metal lines:
\begin{equation}
 \Lambda = \Lambda_{\rm H,He} + \Lambda_{\rm CMB} + \Lambda_{\rm metal}
\end{equation}
The first term $\Lambda_{\rm H,He}$ takes into account Bremsstrahlung and radiative recombination for HII, HeII and HeIII, 
collisional ionization and line cooling for HI, HeI and HeII, and dielectric recombination for HeII. 

In the cooling module of \citet{Shen:2010}, used in many {\sc gasoline} and {\sc gasoline2} simulations, the heating and cooling rates 
due to metals, $H_{\rm metals}$ and $\Lambda_{\rm metals}$, are interpolated from look-up tables constructed from grids of CLOUDY runs 
in the parameter space of density, temperature and redshift. The redshift dependence is due to $UVB=UVB(z)$ used as photoionization 
source in the CLOUDY calculations. \citet{Shen:2010} checked that metal cooling and heating scale linearly with 
the total metallicity $Z$, using simulations where metallicity covers the range $Z\in[3\times10^{\rm -4}Z_{\rm\odot},2Z_{\rm\odot}]$. 
Therefore, CLOUDY models were run for solar and primordial compositions, 
and the differences between the two model grids were tabulated as $H_{\rm metal}(Z_{\rm\odot})$ and $\Lambda_{\rm metal}(Z_{\rm\odot})$. 
Thus, in the code $H_{\rm metal}(Z) = Z/Z_{\rm\odot}H_{\rm metal}(Z_{\rm\odot})$ and 
$\Lambda_{\rm metal}(Z) = Z/Z_{\rm\odot}\Lambda_{\rm metal}(Z_{\rm\odot})$. 
It is also important to note that this cooling model has one discrepancy coming from the fact that the 
primordial cooling is computed with a non-equilibrium solver on-the-fly, while the metal line cooling is done with CLOUDY, 
which is an equilibrium solver. In practical terms, this means that we neglect the contribution of metals to the electron number 
density in the second part of Equation~\ref{eq_cons}. A fully self consistent calculation of the cooling rates would, 
however, imply solving a much more complicated chemical network, which is not yet feasible in statistical samples of zoom-in 
cosmological simulations due to the large computational costs involved. E.g. \citet{Richings:2016} 
used high resolution simulations of isolated galaxies to study the effects of non-equilibrium vs equilibrium chemical networks and 
cooling, and found no significant effect on gas temperatures and surface densities.  However, the same authors show that the ISM 
is much more sensitive to the strength of the UV field and the gas metallicity.

In this work, given that we consider local sources of photoheating and photoionization on top of the metagalactic background, 
the heating and cooling rates due to metals are dependent not only on density, temperature and redshift, but also on the radiation
fields produced by the various types of local sources we consider.

\section{The Local Photoionization Feedback (LPF)}
\label{section_lpf}

While on large cosmological scales, the UVB can be approximated as homogeneous, on galaxy scales the high energy photon sources 
(stars, hot gas, black holes) are distributed highly non-uniformly.  
In zoom-in cosmological simulations, where only a small region of the universe is modeled with high resolution, it is worthwhile 
and computationally feasible under certain assumptions to explore the effects of the non-uniform galactic UV source distribution 
on top of a metagalactic homogeneous UVB. Therefore, in this study we explore the LPF effects on zoom-in galaxies, as well as the 
effects induced by assuming two different UVB models, namely HM12 and FG09.  

For this study we have implemented the LPF model of \citet{Kannan:2016} in the simulation code {\sc gasoline2}. 
The LPF model of \citet{Kannan:2016} has been initially implemented in the moving mesh {\sc arepo} code \citep{Springel:2010}, and 
used to study the local photoionization field effects on galaxy cluster scales in a few idealized simulations.  
This model takes into account three different types of sources producing high-energy ionizing photons, 
apart from the z-dependent UVB of \citet{Faucher:2009}: i) young stars and SNe remnants, ii) post-AGB stars, 
and iii) hot gaseous haloes of galaxies and clusters. All calculations have been done in the optically 
thin approximation. Absorption effects on the spot have been modeled through source dependent escape fractions. 
The first version of this model \citep{Kannan:2014} takes into account only the sources of the first two types. 

\begin{figure}
 \begin{center}
\includegraphics[width=0.50\textwidth]{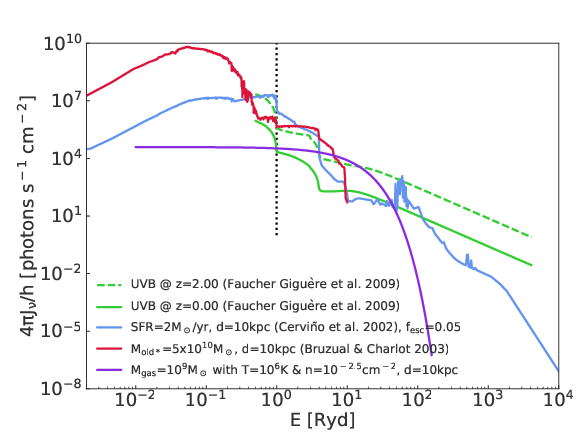}
 \caption{Example of spectra corresponding to the various radiation fields considered within the LPF model.}  
\label{fig_spectra}
 \end{center}
\end{figure}

The first type of source has a spectral energy distribution (SED) made up of two components: a blackbody 
representing the UV contribution from young, massive and hot stars, and  a composite Raymond-Smith hot-plasma 
law describing the contribution of SNe remnants in the soft X-rays \citep{Cervino:2002}. This model has been 
calibrated to match the scaling between star formation rate (SFR) and soft X-rays \citep[e.g.][]{Ranalli:2003}. 
Given that UV and soft X-rays photons have short mean free paths, the escape fraction from the region where this
radiation is produced has been fixed to 5 per cent ($f_{\rm esc,SFR}=0.05$). 
We assume this value for the on-source escape fraction following \citet{Kannan:2014}.
All stellar particles with an age $\rm<10$Myr are assumed to have the same SED. 
Therefore, in the optically thin approximation, the normalization for the radiation field from all 
star forming regions at the position of a gas particle is computed as a direct sum over all the young stellar particles $k$ 
in the simulation, normalized by the distance squared between each source at position $\bm{r_{\rm k}}$ and the gas particle at $\bm{r}$:
\begin{equation}
 \phi_{\rm SFR}(\bm{r}) = f_{\rm esc,SFR} \frac{1}{10^{\rm 7} yr} \Sigma_{\rm k} \frac{M_{\rm k}(age < 10)}{|\bm{r}-\bm{r_{\rm k}}|^{\rm 2}},
 \label{eq_phins}
\end{equation}
where masses are measured in $\rm M_{\rm\odot}$, distances in kpc and ages in Myr.
Figure~\ref{fig_spectra} shows in blue the SED of \citet{Cervino:2002} normalized to a star formation rate of 
2~M$_{\rm\odot}$yr$^{\rm -1}$, as seen from a 10~kpc distance and assuming an escape fraction of the intrinsic radiation 
of 5 per cent. For comparison, the solid and dashed green curves give the UVB SED of \citet{Faucher:2009} at $z=0$ and $z=2$, 
respectively.

In a similar manner to the radiation field from star forming regions, all stellar particles older than 200 Myr 
are assumed to have the same post-AGB SEDs. For this second type of source, \citet{Kannan:2016} uses the SED 
of a 2 Gyr old single stellar population from \citet{Bruzual:2003}, noting that the SED varies only in normalization 
while keeping roughly the same shape for stars with ages from 200 Myr to 13 Gyr. Therefore, all stellar particles $k$
with ages larger than 200 Myr will contribute to the radiation from post-AGB stars at each gas particle position
according to their mass and distance from the gas particle:

\begin{equation}
 \phi_{\rm os}(\bm{r}) = f_{\rm esc,os} \Sigma_{\rm k} \frac{M_{\rm k}(age > 200)}{|\bm{r}-\bm{r_{\rm k}}|^{\rm 2}}.
 \label{eq_phios}
\end{equation}
In this case, the escape fraction of ionizing photons is assumed to be 100 per cent 
($f_{\rm esc,os}=1.0$), 
given that old stars are expected to either have moved from or already dispersed the dense gas at their birth locations 
\citep{Kannan:2014}. 
The red curve in Figure~\ref{fig_spectra} shows the SED of an 2~Gyr old 5$\rm\times$10$^{\rm 10}$M$_{\rm\odot}$ 
single stellar population \citep{Bruzual:2003}, as seen from a 10~kpc distance.

The last type of ionization source that is taken into account in the LPF model is X-ray emission from hot gas. 
Given that in observations most of the hot gas emission is associated with galaxy groups and clusters 
\citep[e.g.][]{Anderson:2015}, \citet{Kannan:2016} have developed the LPF model taking into account the contribution 
only from gas with temperatures $T > 10^{\rm 5.5} K$, and used it to study galaxies in idealized cluster environments. 
However, there are few observations showing that massive spiral galaxies also show emission in X-rays from their haloes 
\citep{Bogdan:2013a,Bogdan:2013b}. Here we apply the model of \citet{Kannan:2016} to zoom-in simulations of L$^{\rm *}$ galaxies, 
for which the virial temperatures are lower than the $T > 10^{\rm 6.5} K$ characteristic of group and cluster environments, but 
we intent to refine the model to include gas at lower temperatures in a further study. 

In the case of the hot gas emission, the temperature sets the SED shape (hardness of the spectra), while the density sets the flux normalization.
Therefore, the emitted spectra varies both in shape and normalization depending on the detailed phase structure of the hot gas, 
and should be quantified as a sum over an infinite number of componets. To bypass this problem, \citet{Kannan:2016} use a three temperature 
bins approximation, centered on 10$^{\rm 6}$, 10$^{\rm 7}$ and 10$^{\rm 8}$~K and with widths of 1.0~dex. 
If a gas particle has the temperature
within one of the bins, its contribution to the radiation field will be computed as it would have the center temperature of the respective bin. 
For each of the three temperature bins $t$, with $t\in[6,7,8]$, the flux from all hot gas particles $k$ with temperatures $T_{\rm k}$ satisfying 
$t-0.5<\log(T_{\rm k}/{\rm K})<t+0.5$ at the position $\bm{r}$ is computed as:
\begin{equation}
 F_{\rm\nu}(\bm{r}) = \varLambda_{\rm\nu}(10^{\rm t})\frac{n_{\rm e}/n_{\rm H}}{(n_{\rm e}/n_{\rm H}+n_{\rm (i)}/n_{\rm H})}\frac{1}{4\pi}\phi_{Tt}(\bm{r})
\end{equation}
and the contribution of all relevant particles $k$ to the radiation field in the bin $t$ at position $\bm{r}$ is:
\begin{equation}
 \phi_{Tt}(\bm{r}) = \Sigma_{\rm k} \frac{\rho_{\rm k}m_{\rm k}}{\mu_{\rm k}^{\rm 2}m_{\rm p}^{\rm 2}|\bm{r}-\bm{r_{\rm k}}|^{\rm 2}}
 \label{eq_phiT}
\end{equation}
where $\rho_{\rm k}$, $m_{\rm k}$, $\bm{r_{\rm k}}$ and $\mu_{\rm k}$ are the density, mass, position and mean molecular weight of particle $k$, 
and $m_{\rm p}$ is the proton mass. The temperature function $\varLambda_{\rm\nu}(10^{\rm t})$ has the characteristic form for the thermal free-free
emission:
\begin{equation}
 \varLambda_{\rm\nu}(T) = 6.8\times10^{\rm -38} T^{\rm -1/2} e^{\rm h\nu/k_{\rm B}T}Z^{\rm 2}g_{\rm ff}
\end{equation}
where $k_{\rm B}$ is the Boltzmann constant, $g_{\rm ff}$ is the free-free gaunt factor, and $Z$ in this case is the mean ionic charge. 
The flux from hot gas with temperature of 10$^{\rm 6}$K (hence in the first of the temperature bins considered), 
densities of 10$^{\rm -2.5}$cm$^{\rm -2}$ and primordial composition, as seen from 10~kpc is shown as the violet curve in 
Figure~\ref{fig_spectra}.

The five types of local radiation field at each gas particle position are used in the Equation~\ref{phot_rates} for the photoionization rate 
of primordial ion species $(i)$ as: 
\begin{equation}
 \varGamma_{\rm(i)} = \varGamma_{\rm(i)UVB} + \phi_{\rm SFR}\gamma_{\rm(i)SFR} + \phi_{\rm os}\gamma_{\rm(i)os} + \Sigma_{\rm t\in[6,7,8]}\phi_{\rm Tt}\gamma_{\rm(i)Tt}
\end{equation}
where the $\gamma_{\rm(i)}$ terms are the coefficients that ensure the correct units of photoionization rate of s$^{\rm -1}$. 
In a similar way, the photoheating for each primordial species (i) that enters in Equation~\ref{heating_eq} is the 
sum of the six different contributions: one from the UVB and the other five from the local radiation fields: 
\begin{equation}
 \epsilon_{\rm(i)} = \epsilon_{\rm(i)UVB} + \phi_{\rm SFR}\bar{\epsilon}_{\rm(i)SFR} + \phi_{\rm os}\bar{\epsilon}_{\rm(i)os} + \Sigma_{\rm t\in[6,7,8]}\phi_{\rm Tt}\bar{\epsilon}_{\rm(i)Tt}
\end{equation}
where the $\bar{\epsilon}$ factors give the correct units for photoheating of erg~cm$^{\rm 3}$~s$^{\rm -1}$.

The cooling and heating rates due to metals, $\Lambda_{\rm metal}$ and $H_{\rm metal}$, 
are interpolated from the table of \citet{Kannan:2016}, 
which was constructed from CLOUDY models run on an eight dimensional parameter space
$(z,\rho,T,\phi_{\rm SFR},\phi_{\rm os},\phi_{\rm T6},\phi_{\rm T7},\phi_{\rm T8})$. 
The redshift parameter encapsulates the z-dependent 
UVB of FG09, and the metallicity is taken into account by applying the scaling of \citet{Shen:2010} discussed previously.  
Practically, this implied running grids of CLOUDY models for gas densities 
10$^{\rm -9}\leqslant\rho\leqslant$10$^{\rm 4}$cm$^{\rm -3}$ and temperatures 10$^{\rm 2}\leqslant T\rm\leqslant$10$^{\rm 9}$K, 
with spacing of 1.0~dex for $\rho$ and 0.1~dex for T, for a large sample of incident radiation fields 
produced as the sum of the UVB($z$) and the $\phi$-normalized typical spectra shown in Figure~\ref{fig_spectra}. 
The SEDs are constructed by varying  
$\phi_{\rm SFR}$ between 10$^{\rm -5}$ and 10$^{\rm 3}$ M$_{\rm\odot}$yr$^{\rm -1}$kpc$^{\rm -2}$, 
and $\phi_{\rm os}$ between 10$^{\rm 6}$ and 10$^{\rm 12}$ M$_{\rm\odot}$kpc$^{\rm -2}$, 
both with a spacing of 1.0~dex, and $\phi_{\rm T6}$, $\phi_{\rm T7}$ and 
$\phi_{\rm T8}$ in the range [10$^{\rm 17.5}$,10$^{\rm 23.5}$]~cm$^{\rm -5}$ with a 2.0~dex spacing.
Given that galaxies only have hot halos at low redshifts, the cooling table has five dimensions 
($\rho$,$T$,$\phi_{\rm SFR}$,$\phi_{\rm os}$,UVB($z$)) for 3.0$<z<$9.0, and eight dimensions 
($\rho$,$T$,$\phi_{\rm SFR}$,$\phi_{\rm os}$,$\phi_{\rm T6}$,$\phi_{\rm T7}$,$\phi_{\rm T8}$,UVB($z$)) only for $z\leqslant$3.0.
The UVB dimension in the cooling table covers the range 0.0$<z<$9.0 with a 0.5~dex spacing, 
and, for $z>$9.0, the gas is assumed to be purely primordial. 
The table contains the heating and cooling rates due to metals only, computed as differences between the CLOUDY grids 
run for solar metallicity and the equivalent ones run for a primordial gas composition. 
Large grid spacings were used to keep the cooling table to a reasonable size. 
For details on how the grid spacing affects the accuracy of cooling and heating rates see \citet{Kannan:2016}.

We model the shielding of the dense gas from the local radiation field by a simple density dependent relation. 
All gas particles with densities $n>n_{\rm 0}=0.1$cm$^{\rm -3}$ \emph{see} an attentuated local radiation field 
$\phi_{\rm attenuated}=\phi\times exp(1-n/n_{\rm 0})$. 
\citet{Kannan:2016} follow \citet{Ceverino:2010}, and consider all gas with densities 
$n>n_{\rm 0}$=0.1~cm$^{\rm -3}$ to feel no local radiation field. We use the factor $exp(1-n/n_{\rm 0})$ to attenuate the 
local radiation field, with the same density threshold $n_{\rm 0}=0.1cm^{\rm -3}$ as Kannan et al., in order to 
have a smoother transition between no shielded and partially shielded regions. A more physical motivated 
attenuation would be e.g. also an exponential \citep[e.g.][]{Richings:2014,Richings:2016,Hopkins:2018a}, 
but whose argument would depend on the local column density computed via a Sobolev-like approximation \citep{Sobolev:1957,Gnedin:2009}. 
We plan to explore this kind of more realistic attentuations in a future work.

\subsection{LPF in {\sc gasoline2}}
\label{lpf_in_gas2}

{\sc gasoline2} is a modern-SPH code build on top of the N-body parallel code {\tt pkdgrav} \citep{Stadel:2001}. 
The novelty of the code are the SPH expressions for force and internal energy, which use the so-called 
'geometric density average' (GDP). The GDP is a particular type of SPH pressure gradient formulation which 
accurately conserves entropy \citep{Monaghan:1992}, minimizes numerical errors in regions of large density gradients \citep{Ritchie:2001}, 
and also minimizes numerical surface tension effects. In this manner the problem of artificial cold blobs, present in the classical
SPH formulations \citep{Agertz:2007}, and therefore in the older version of the code, is greatly reduced. 
The sub-grid turbulence diffusion is modeled using the trace-free local shear tensor \citep{Wadsley:2008,Shen:2010}, ensuring 
a correct transport of entropy and a physically sound evolution of hydrodynamical instabilities.  
A lower velocity noise and a better treatment of inviscid flows are obtained using the time-dependent artificial viscosity of 
\citet{Cullen:2010}, and spurious viscosity in convergent flows is minimized by a normal-velocity-gradient shock detector.
To ensure that physical quantities are accurately conserved locally, the Kick-Drift-Kick 2nd order leapfrog integration with 
hierarchical powers of two time stepping has been complemented with the time step limiter of \citet{Saitoh:2009} applied pairwise.
In this manner it is ensured that neighboring particles have similar time-steps, and as a result the time integration is more 
closely symplectic. Last but not least, {\sc gasoline2} uses Wendland \citep{Wendland:1995} smoothing kernels with 50 neighbors, 
which improves convergence, as demonstrated by \citet{Dehnen:2012}.

The five types of local photoheating and photoionization sources described in the previous section are computed on the 
gravity tree of {\sc gasoline2} at the same time and in a similar way as the evaluation of the gravity force on each particle.
In this manner, the time needed for the computation of the local field at each gas particle position adds little to the total run time.  

{\sc gasoline2} is build on {\tt pkdgrav}, a parallel N-body code which uses a binary kD tree to construct a
hierarchical representation of the mass distribution, which is subsequently used to compute the gravity forces via multipole expansions. 
The kD tree (where $k=3$) is constructed top-down starting from the root-cell and bisecting recursively the bounding box containing all 
the particles in each one cell. Therefore, each cell but the leaf cells has two child cells. This spatial binary tree structure, where 
the bounds of each cell are saved and the particles are ordered in memory, is used for both gravity calculation and nearest neighbor search. 
Next, the tree is walked bottom-up to calculate all the needed cell properties, e.g. center-of-mass and reduced multipole moments. 
During this step we added the computation of the photoionization/photoheating sources, by directly summing the contributions from the 
relevant particles into source dependent mass terms. For all but the leaf cells, all these quantities can be obtained from the two child cells. 
For the leaf cells the center of mass, reduced multipoles and the source dependent mass terms are trivially computed from their respective particles. 
In a third step, the tree is walked top-down to construct for each particle two types of interaction list: a particle-cell one (c-list)
and a particle-particle one (p-list). These two lists are used to compute the gravitational force and the radiation field at each particle position. 
The c-list contains the multipoles (plus their center of mass and their radiation source dependent mass terms) of the cells that are sufficiently 
far away such that their gravitational interaction with the particle can be approximated by the multipole expansion. In a similar way, 
the c-list also contains the masses and distances needed to compute the radiation field from the farther away sources using Equations~\ref{eq_phins}, 
\ref{eq_phios}, \ref{eq_phiT} where $r_{\rm k}$ now refers to the distance between the particle and the center-of-mass of the cells in the list, 
and $M_{\rm k}$ refers to the respective source dependent masses. The p-list, on the other hand, contains the particles that are close enough that 
their gravitational interaction and their radiation field contribution need to be computed by direct summation. Therefore, for each of the five 
types of radiation sources, the field at each gas particle position is a sum of two terms: one computed from the c-list, and the other from the p-list.

\section{Test cases: NIHAO simulations}
\label{sims}

The NIHAO \citep{Wang:2015} project bridges the gap between cosmological high resolution 
simulation of a certain galaxy type (e.g. Milky Way) and large box simulations that host all types of galaxies, but at lower resolutions, 
by providing a statistical sample of cosmological hydrodynamical zoom-in simulations covering almost three orders of magnitude in halo mass 
(from $\rm\sim4\times10^{\rm 9}M_{\rm\odot}$ to $\rm\sim4\times10^{\rm 12}M_{\rm\odot}$).
The code used to simulate the NIHAO galaxies is the improved version of {\sc gasoline2} \citep[][see Section~\ref{lpf_in_gas2}]{Wadsley:2017}.
The dark matter haloes have been selected from the dark matter only simulations of \citet{Dutton:2014} which use the $\rm\Lambda$CDM cosmology 
of the \citet{Planck:2014}. The number of dark matter particles has been chosen to be approximately constant ($\rm\sim$10$^{\rm 6}$) 
accross all the halo mass range probed, to ensure that mass profiles are resolved on stellar half mass radius scales.

In the standard implementation of the {\sc gasoline2} code used to run NIHAO, the only source of photoheating and 
photoionization is the UVB of HM12, which was also used to compute the CLOUDY models needed for the heating and 
cooling rates due to metals. Following \citet{Shen:2010}, gas cools through hydrogen, helium and metal lines, and through Compton 
scattering of the CMB, as explained in Section~\ref{heat_cool}. Star formation follows a Kennicutt-Schmidt relation, 
where the gas eligible has an upper temperature limit of 15000~K and a lower density limit of 10.3~cm$^{\rm-3}$. 
For the NIHAO set-up two types of stellar feedback are considered:
pre-heating of gas by the massive progenitors of SNe II also known as `early stellar feedback' \citep{Stinson:2013a}, 
and blast-waves produced by SNe II  \citep{Stinson:2006}. \citet{Stinson:2013a} chose the feedback efficiency parameters   
such that one Milky Way mass galaxy respects the abundance matching relation at all redshifts  
\citep{Behroozi:2013,Moster:2013,Kravtsov:2014}. This choice of feedback parameters has been shown to make the NIHAO galaxy 
sample follow the M$_{\rm star}$--M$_{\rm halo}$ relation at all redshifts and over the entire halo mass range probed \citep{Wang:2015}. 
For the stellar evolution, {\sc gasoline2} uses  the Initial Mass Function (IMF) of Chabrier \citep{Chabrier:2003}, 
and the heavy elements enrichment is computed following the SNe Ia yields of \citet{Thielemann:1986} and the SNe II of \citet{Woosley:1995}.

There is a growing number of papers that have tackled various topics in galaxy formation using the NIHAO simulations.
\citet{Obreja:2016,Obreja:2018} e.g. have shown that the stellar dynamics of the various NIHAO galaxy components 
follow a plethora of observations in terms of rotational support, shapes, sizes and angular momentum content.  
\citet{Buck:2017} have post-processed NIHAO simulations with the radiation transfer code {\sc grasil-3d} 
\citep{Dominguez-Tenreiro:2014} to generate Hubble Space Telescope images and study the nature of clumpy galaxies at high redshift. 
\citet{Maccio:2016} and \citet{Dutton:2019} have shown that when computed properly in simulations, by constructing HI line profiles 
and taking into account the effect of galaxy inclination, the HI velocity function in the nearby universe is in very good agreement 
with observational data. Recently, \citet{Buck:2019} have resimulated at higher resolutions four of the more massive galaxies in the 
original sample, to study the properties of the dwarfs around Milky Way and Andromeda galaxies.

\begin{table}
\centering
\begin{tabular}{ccccc}
\hline
Sim & $\rm\epsilon_{\rm gas}$ [kpc] & $\rm m_{\rm gas}$ [M$_{\rm\odot}$] & $\rm\epsilon_{\rm dark}$ [kpc] & $\rm m_{\rm dark}$ [M$_{\rm\odot}$]\\
\hline
g1.08e11 & 0.2 & 4.0$\rm\times$10$^{\rm 4}$ & 0.5 & 2.2$\rm\times$10$^{\rm 5}$\\
g8.26e11 & 0.4 & 3.2$\rm\times$10$^{\rm 5}$ & 0.9 & 1.7$\rm\times$10$^{\rm 6}$\\
g2.79e12 & 0.4 & 3.2$\rm\times$10$^{\rm 5}$ & 0.9 & 1.7$\rm\times$10$^{\rm 6}$\\
\hline
\end{tabular}
\centering\caption{The dark matter and gas particle masses, and the hydro and gravitational force softenings for the three NIHAO simulations.}
\label{table_runs}
\end{table}

In the light of all the previous works using the NIHAO sample that have shown good agreement with observational data 
we resimulated with the new version of the code three NIHAO galaxies, to test the effects of the LPF implementation: 
the dwarf g1.08e11, the Milky Way analogue g8.26e11, and the most massive object in the sample g2.79e12. 
The simulations set up in terms of spatial and mass resolution for both gas and dark matter particles are given in Table~\ref{table_runs}.
Therefore, for each of the three galaxies, we have three different runs: 
\begin{itemize}
 \item the standard NIHAO which uses a HM12 UVB, shown in grey color through the figures of this paper and denoted by HM12,
 \item the run which uses the FG09 UVB instead of the HM12, shown in blue and named FG09,
 \item the LPF run which uses the local photoionization sources implementation described in Section~\ref{section_lpf} on top of the FG09 UVB, shown in red and named FG09+LPF.
\end{itemize}

For each galaxy, the simulations have been done from the same initial conditions and with the same resolution.
In the following sections we discuss in detail the differences between these three models in terms of galaxy properties at $z=0$ and formation histories,  and finally make some comparison with observational data. 

\begin{figure}
\begin{center}
 \includegraphics[width=0.55\textwidth]{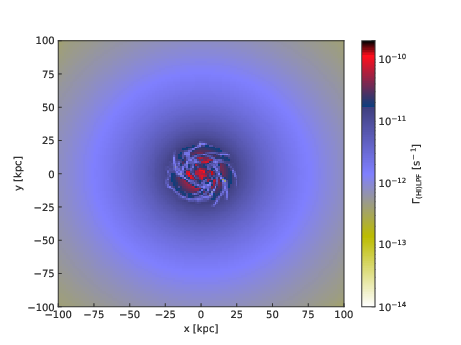}\\
 \includegraphics[width=0.47\textwidth]{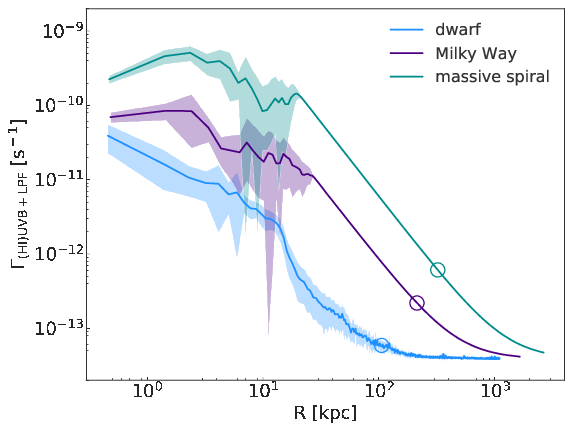}\\
 \includegraphics[width=0.47\textwidth]{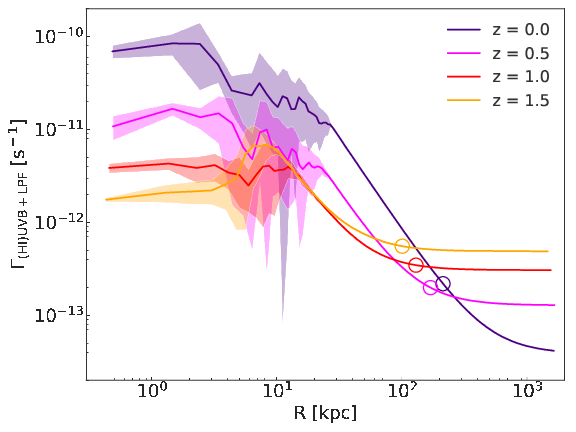}\\
 \caption{\textbf{Top:} face-on map of the LPF contribution to the HI photoionization rate, 
 $\rm\Gamma_{\rm(HI)LPF}$, for the Milky Way like galaxy g8.26e11 at $z=0$. At this redshift, the FG09 model predicts a rate of 
 $\rm\Gamma_{\rm(HI)UVB}\backsimeq4\times10^{\rm -14}s^{\rm -1}$. 
 \textbf{Center:} mean radial profiles of $\rm\Gamma_{\rm(HI)}$ in the plane of the disc for the three simulated 
 galaxies. The shaded regions mark the standard deviation, while the open dots give the corresponding virial radii (see Table~\ref{table_z0properties} for exact values). \textbf{Bottom:} The $z$-evolution of the $\rm\Gamma_{\rm(HI)}$ profile 
 for the Milky Way like galaxy. The open dots mark the corresponding virial radii at each redshift.}  
\label{fig_photoionization_map}
\end{center}
\end{figure}

\begin{figure*}
\begin{center}

 \textbf{\underline{g1.08e11}}\hspace*{4.5cm}\textbf{\underline{g8.26e11}}\hspace*{4.5cm}\textbf{\underline{g2.79e12}}\\

 \includegraphics[width=0.31\textwidth]{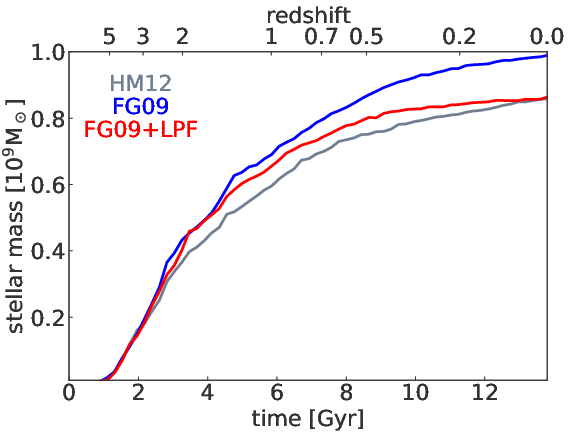}
 \includegraphics[width=0.31\textwidth]{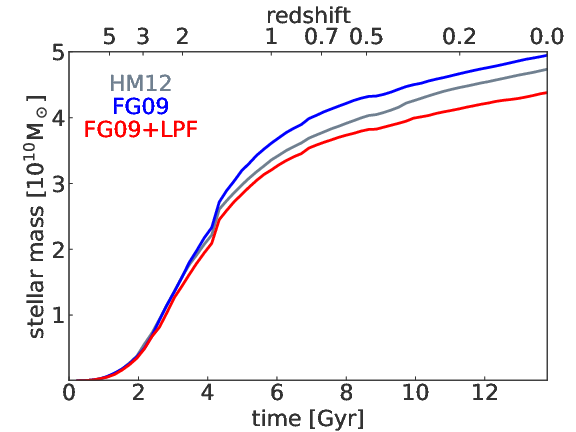}
 \includegraphics[width=0.31\textwidth]{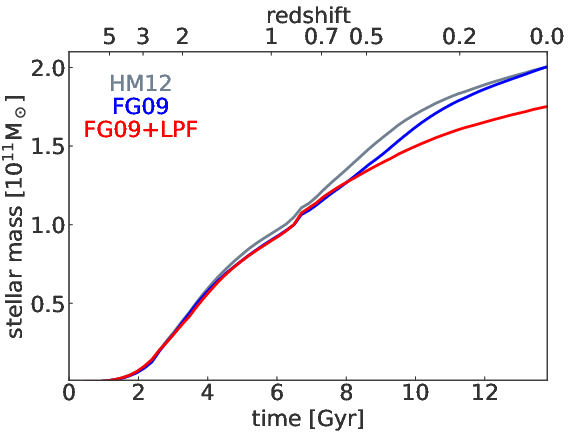}\\

 \includegraphics[width=0.31\textwidth]{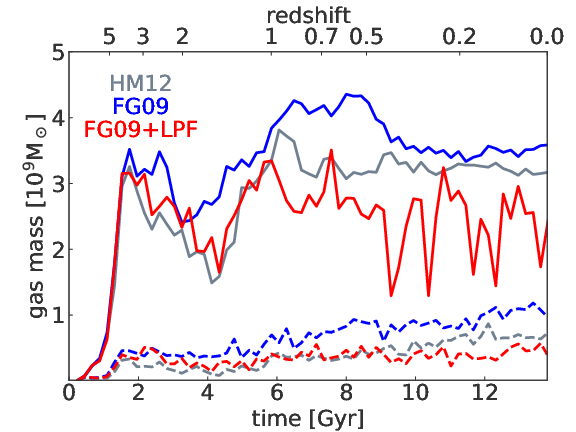}
 \includegraphics[width=0.31\textwidth]{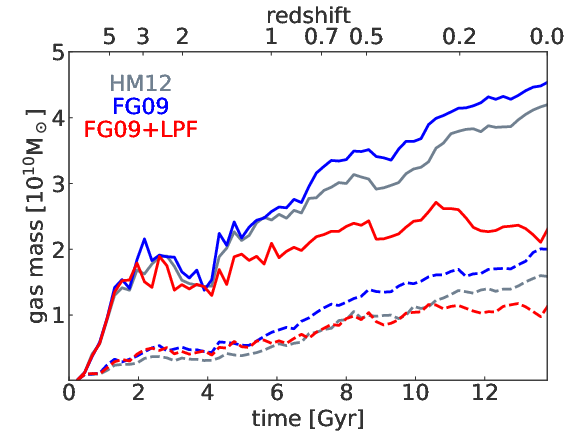}
 \includegraphics[width=0.31\textwidth]{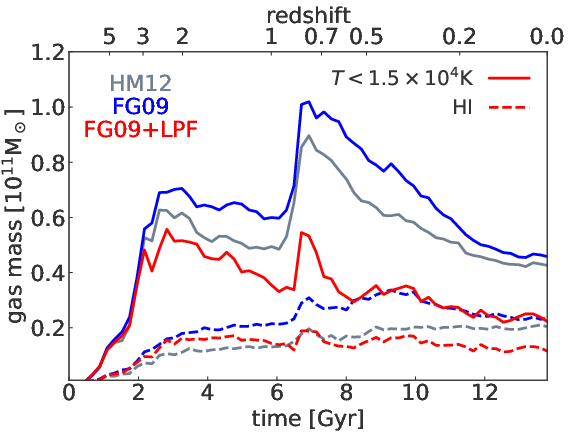}\\

 \includegraphics[width=0.31\textwidth]{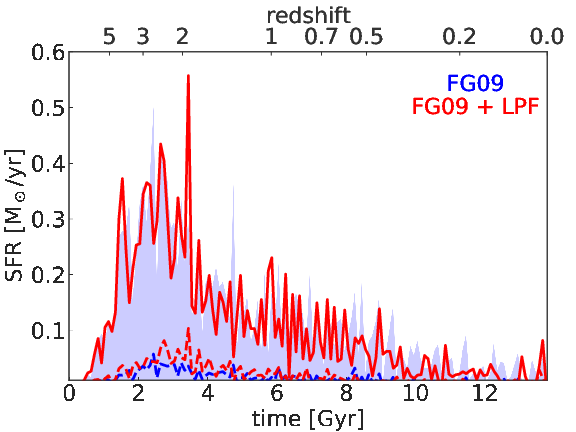}
 \includegraphics[width=0.31\textwidth]{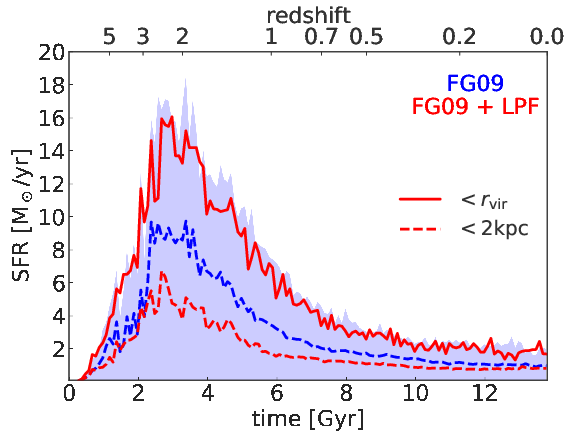}
 \includegraphics[width=0.31\textwidth]{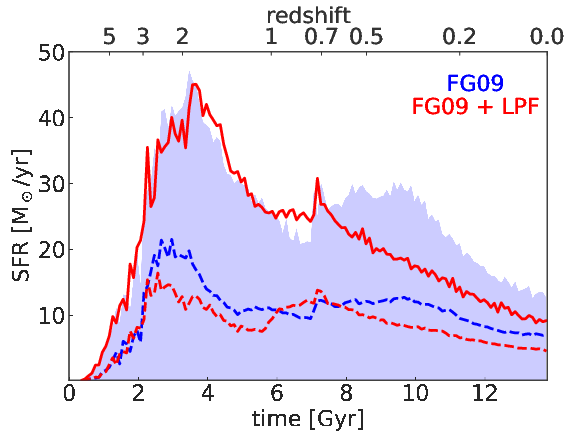}\\

 \caption{Comparison between the evolution of stellar (top) and gas (center) mass, and star formation rates (bottom) for the galaxies g1.08e11 (left), g8.26e11 (center) and g2.79e12 (right) simulated with the HM12 (grey), FG09 (blue) and FG09+LPF (red) models.
 The solid and dashed lines in the central panels refer to cold ($T<$15000~K) and HI gas, respectively. 
 In the bottom panels, the shaded blue areas give the star formation rates within $r_{\rm vir}$ for the FG09 model, while the 
 solid red curves the corresponding SFR for FG09+LPF. The dashed curves in these panels represent the SFR evolution within a sphere 
 of fixed physical radius of 2~kpc.}
\label{mass_tracks}
\end{center}
\end{figure*}

\begin{figure*}
\begin{center}
 \includegraphics[width=0.31\textwidth]{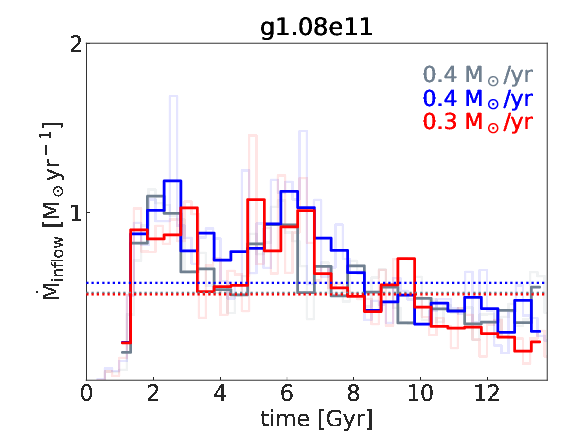}
 \includegraphics[width=0.31\textwidth]{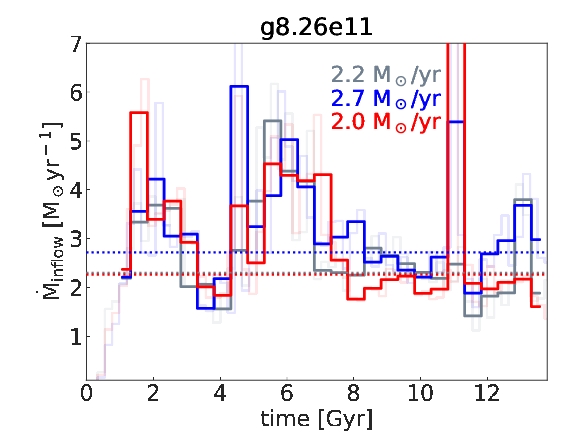}
 \includegraphics[width=0.31\textwidth]{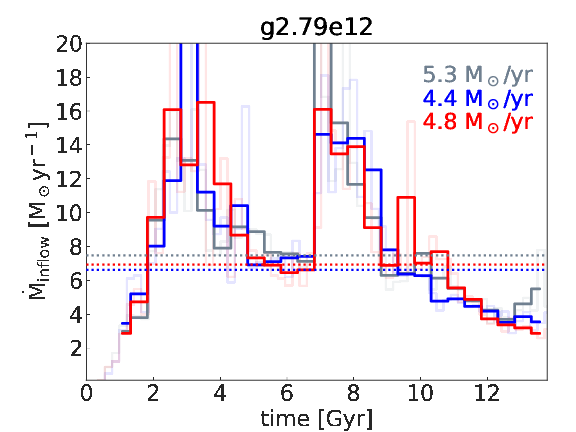}\\
 \includegraphics[width=0.31\textwidth]{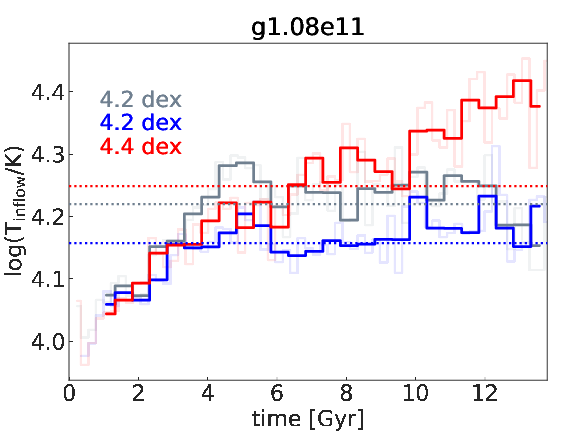}
 \includegraphics[width=0.31\textwidth]{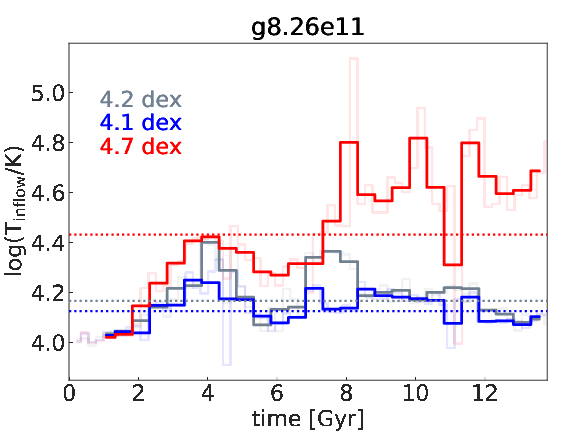}
 \includegraphics[width=0.31\textwidth]{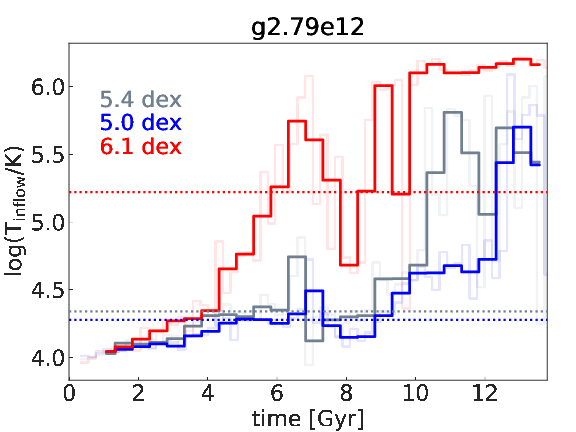}\\

 \caption{Evolution of the inflow rates (top) and of the median temperatures of inflow gas (bottom) 
 for the galaxies g1.08e11 (left), g8.26e11 (center) and g2.79e12 (right) run using the HM12 (grey), FG09 (blue) and FG09+LPF (red) models. The faint cuves in the background are the rates and temperatures measured on 100~Myr time scale, 
 while the solid colors represent averages over 500~Myr.
 The dotted horizontal lines show the 50 per cent quantiles of the corresponding inflow rates and median temperatures distributions, respectively.
 The values for the 50 per cent quantiles are given in Table~\ref{table_z0properties}. The numbers in each 
 panel give the respective average rates and temperatures for the last 4~Gyr. In the case of g8.26e11, the data point corresponding 
 to the minor merger at time$\rm\sim$11~Gyr has been excluded when computing these averages.}
\label{fig_accretion}
\end{center}
\end{figure*}

\begin{figure*}
\begin{center}
 \includegraphics[width=0.31\textwidth]{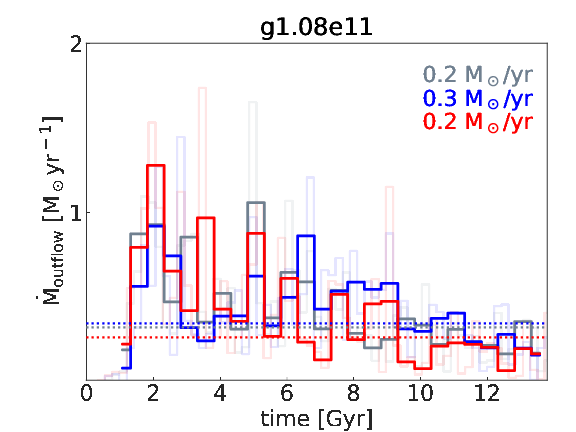}
 \includegraphics[width=0.31\textwidth]{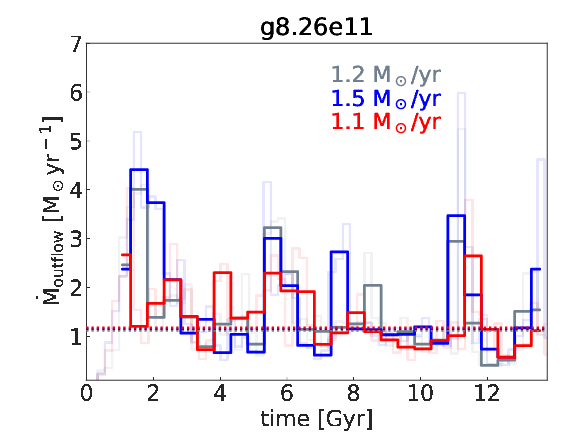}
 \includegraphics[width=0.31\textwidth]{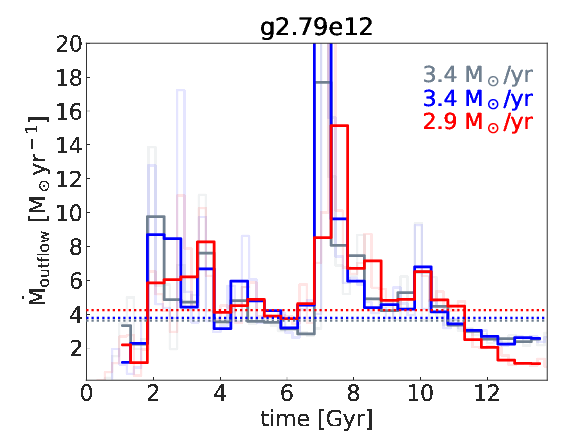}\\
 \includegraphics[width=0.31\textwidth]{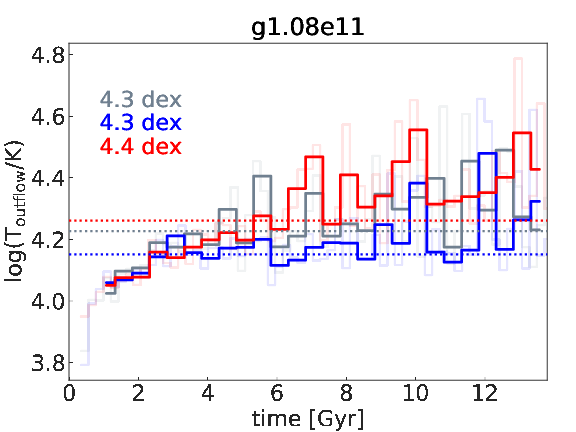}
 \includegraphics[width=0.31\textwidth]{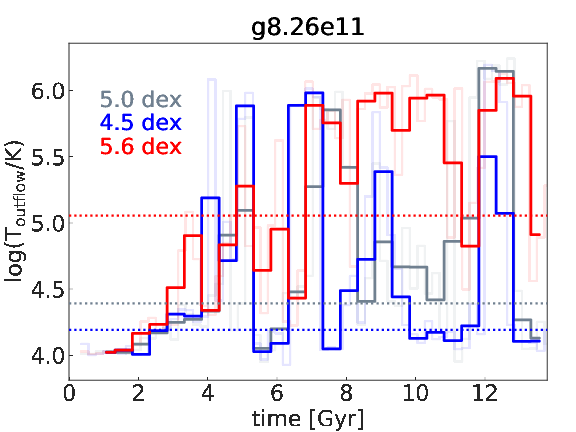}
 \includegraphics[width=0.31\textwidth]{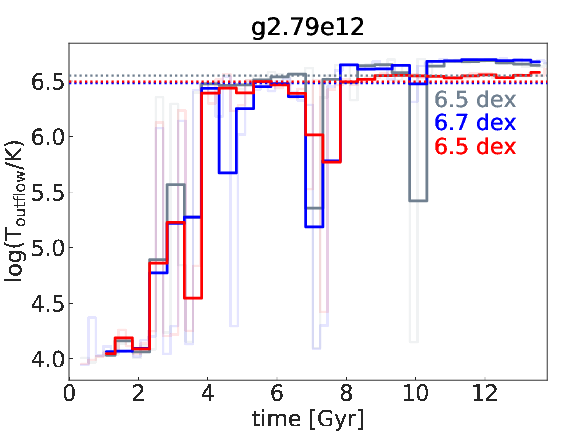}\\

 \caption{Evolution of the outflow rates (top) and of the median temperatures of outflow gas (bottom) 
 for the galaxies g1.08e11 (left), g8.26e11 (center) and g2.79e12 (right) run using the HM12 (grey), FG09 (blue) and FG09+LPF (red) models. The faint cuves in the background are the rates and temperatures measured on 100~Myr time scale, 
 while the solid colors represent averages over 500~Myr. The dotted horizontal lines show the 50 per cent quantiles of the corresponding inflow rates and median temperatures distributions, respectively.
 The values for the 50 per cent quantiles are given in Table~\ref{table_z0properties}. 
 The numbers in each panel give the respective average rates and temperatures for the last 4~Gyr.}
\label{fig_outflow}
\end{center}
\end{figure*}

\section{Results}
\label{results}

One of the most important physical quantities governing the phase structure of gas in 
galaxies is the hydrogen photoionization rate $\rm\Gamma_{\rm(HI)}$. The radiation field model 
discussed in Sections~\ref{heat_cool} and \ref{section_lpf} naturally results in an inhomogenous and enhanced 
$\rm\Gamma_{\rm(HI)}$ with respect to the UVB on galaxy scales. 
To illustrate how $\rm\Gamma_{\rm(HI)}$ varies accross our 
simulated galaxies, we show in Figure~\ref{fig_photoionization_map}
the face-on map of the LPF contribution to the HI photoionization rate for the Milky Way-like object 
g8.26e11 at $z=0$ (top panel), the mean radial profiles of $\rm\Gamma_{\rm(HI)}$ in the plane of 
the disc for all three simulated galaxies (middle), and the evolution of the $\rm\Gamma_{\rm(HI)}$
profile with redshift for g8.26e11 (bottom). 

The map was constructed by taking a slice of 
10~kpc in face-on projection, centered on the equatorial plane of the galaxy, and averaging 
$\rm\Gamma_{\rm(HI)LPF}$ along the line-of-sight. It is interesting to note the large difference in 
$\rm\Gamma_{\rm(HI)LPF}$ between the spiral arms of the galaxy and the inter-arms and very inner regions. 
From the mean radial profiles at $z=0$ (middle panel of Figure~\ref{fig_photoionization_map}), 
it is clear that $\rm\Gamma_{\rm(HI)}$ in the galaxy region is orders of magnitude 
higher than the UVB level, which was assumed to be $\rm\Gamma_{\rm(HI)UVB}\backsimeq4\times10^{\rm -14}s^{\rm -1}$ (FG09). 
We should mention that $\Gamma_{\rm HI}$ computed by FG assumes UV escape fraction
from galaxies that is different from 5\% assumed here. So if one does self consistent
models the value for the back ground radiation will be higher. But this will not be much
higher than $10^{\rm -13}$.
Also, the radial position where $\rm\Gamma_{\rm(HI)}$ saturates to the background 
level moves to larger $R$ as the galaxy mass increases, as illustrated by the open circles marking the virial 
radius of each galaxy. 
Between the galaxy regions ($R\lesssim0.2r_{\rm vir}$) and the corresponding 
saturation radii, $\rm\Gamma_{\rm(HI)}$ is well described by a fixed power law: $\rm\Gamma_{\rm(HI)}(R)\propto R^{\rm -2}$, 
whose origin is the assumption of the inverse squared distance factor in the local photoionization field.
This type of profile is a prediction for $z=0$ of our particular local radiation field 
model, which could be tested in the future on combined H$_{\rm\alpha}$ and HI integral field unit obsevational 
data on nearby galaxies, as suggested by the analysis of e.g. \citet{Fumagalli:2017}.

The bottom panel of Figure~\ref{fig_photoionization_map} shows how the radial profile of $\rm\Gamma_{\rm(HI)}$ 
varies with $z$ for the Milky Way-like galaxy, and provides a clear indication on how the local radiation field 
depends on the star formation rate. E.g. at $z=1.5$ the SFR of this galaxy is about a factor of 5 higher than at 
$z=0$ ($\rm\sim$10M$_{\rm\odot}$yr$^{\rm -1}$ vs $\rm\sim$2M$_{\rm\odot}$yr$^{\rm -1}$), and concentrated in the 
disc region (3$\lesssim$R$\lesssim$20~kpc), giving rise to a peak in $\rm\Gamma_{\rm(HI)}$ at these radii. Also, 
from this figure one can appreciate that at redshifts closer to the peak of the cosmic star formation rate density
and AGN activity, e.g. $z=1.5$, where the UVB level is significantly higher than at $z=0$, 
the photoionization rate inside the galaxy is between a factor of 2 and 10 larger that
the level of the background. By comparison, the difference between $\rm\Gamma_{\rm(HI)}$ 
in the galaxy region and in the IGM is much higher at $z=0$ (up to 3 orders of magnitude).

The Milky Way-like galaxy g8.26e11 also provides us with a good case to test if 
our LPF model predicts a reasonable interstellar radiation field (ISRF). This galaxy in particular has been 
shown to be a good Galaxy analogue, not only in terms of stellar mass and SFR, but also in terms of thin + thick 
disc structure \citep{Obreja:2018a}. The ISRF in both external galaxies and the Milky Way is not a direct observable, 
and in extragalactic objects computing it is particularly difficult because of dust obscuration effects \citep[e.g.][]{DeLooze:2012}. 
The Galaxy, and especially the solar neighborhood have received much more attention in this respect 
\citep[e.g.][]{Habing:1968,Draine:1978,Gondhalekar:1980,vanDishoeck:1994}.
Therefore, we used the average local radiation field at $R$=8~kpc in the plane of the disc (-0.5$<z<$0.5~kpc) of g8.26e11, 
to compute the ISRF at this position by summing the normalized spectra shown in Figure~\ref{fig_spectra} for the UVB, 
star forming regions, old stars and hot gas (10$^{\rm 5.5}<T<$10$^{\rm 6.5}$~K). The contribution from gas with 
temperatures $>$10$^{\rm 6.5}$K to the total ISRF is negligible. What we obtained is that the averaged flux in the 
UV wavelength range 0.1$<\lambda<$0.2~$\rm\mu$m is 0.028$\rm\pm$0.005~eV/cm$^{\rm 3}$. By comparison, the observational 
data for the solar neighborhood ISRF by \citet{Witt:1973} and \citet{Henry:1980} in the same wavelength range resulted 
in values of 0.041$\rm\pm$0.001 and 0.050$\rm\pm$0.006~eV/cm$^{\rm 3}$, respectively. Considering that g8.26e11 is not 
an isolated galaxy model \emph{made} to resemble the Milky Way, the factor of  less than 2 between the UV ISRF of the simulated 
galaxy and the Milky Way observations is quite resonable given the variation in the observational estimates \citep{vanDishoeck:1994}. 
It would be nevertheless interesting to test in the future this particular LPF prediction for a dynamical + stellar population model 
of the Milky Way \citep[e.g.][]{Popescu:2017} coupled with a realistic hot halo.

\subsection{Formation histories}
\label{formation}

We first look at how the stellar mass grows through cosmic time in the HM12, FG09 and FG09+LPF models in the top panels 
of Figure~\ref{mass_tracks}. For all three galaxies, the LPF (red curves) slows down the growth of stellar mass after $z\simeq1.5$ 
compared with the runs with only UV backgrounds (grey and blue curves). 
Given that two of the three photoionization sources that we include in the model (old stars and the hot haloes of galaxies) build up in time, 
we expect the LPF to be most effective in shaping the galaxy evolution at lower redshifts ($z<1.5$). 
At $z=0$ the stellar masses in the FG09+LPF cases are $\rm\sim$20 per cent smaller than in the corresponding FG09 ones. 
This fractional decrease in $z=0$ stellar masses is less than the 
$\rm\sim30$ per cent found by \citet{Kannan:2014}. 
While we see similar effects of the local radiation field in our dwarf galaxy with 
respect to the equivalent one of \citet{Hopkins:2018a}, LPF is also effective in decreasing 
the stellar mass in more massive galaxies, where the LEBRON scheme of Hopkins et al. is not. We think this 
can be explained by a combination of differences in the star formation models, self-shielding and resolution. 
The higher resolutions of Hopkins et al. allow the use of a high SF density threshold of 10$^{\rm 3}$cm$^{\rm -3}$, in comparison 
with ours which is two orders of magnitude smaller (10~cm$^{\rm -3}$). Apart from this density condition, we only use a temperature
cut of 15000~K, while Hopkins et al have three additional criteria: gas has to be self-graviting, thermally Jeans unstable and 
self-shielded according to the condition of \citet{Krumholz:2011}. These four conditions of Hopkins et al. coupled with the 
fact that the local radiation field has little to no effect on gas with densities $>10^{\rm -2}$cm$^{\rm -3}$, most likely result 
in very small differences in the gas mass eligible to form stars in higher mass galaxies (see fig 6 of Hopkins et al.).  
In our simulations there is also a two order of magnitude difference between the density at which the local radiation field starts being attenuated ($n>$10$^{\rm -1}$cm$^{\rm -3}$) and where gas is able to transform into stars ($n>$10~cm$^{\rm -3}$). 
However, we find that LPF results in a median 20\% less gas eligible to form stars after $z=2$ when compared with the 
FG09 model. In our case, this is due to the change in the gas temperature distribution towards higher values (see Section~\ref{z0_gas}), 
and not so much to a change in the density distribution.

All final stellar masses are given in Table~\ref{table_z0properties}. When comparing the UVB only runs, 
it can be noted that the weaker FG09 model 
results in more stars than the HM12 one in the dwarf galaxy at $z=0$. However, for the Milky Way galaxy, the differences in the stellar 
mass growth between the FG09 and HM12 models are significantly smaller than in the dwarf case, and the two models become basically indistinguishable for the more massive object g2.79e12. This result suggests that for galaxies above the Milky Way mass range, 
the assumption of a particular homogeneous and $z$-dependent UVB model is not likely to influence the final stellar mass, but it will 
impact significantly the galaxies in the dwarf regime. An interesting fact is that for the dwarf galaxy, the stronger UVB (HM12) mimicks 
the stellar mass growth in the case of local photoionization field on top of a weaker UVB.

The central panels of Figure~\ref{mass_tracks} show with solid lines the cold gas mass evolution, 
where by cold we understand gas at temperatures below the star formation threshold of 15000~K. 
Here, the effect of LPF is more marked, especially for the two more massive galaxies for which the 
cold gas mass in the virial radius is approximately half that of the corresponding FG09 run at lower $z$s 
(solid red vs solid blue curves). 
The LPF runs result in almost overlapping tracks for the cold gas mass within the virial 
radius and within the galaxy region (not shown in this figure). At lower redshifts, 
and especially for the two more massive galaxies, LPF is effective in keeping warm/hot virtually all the gas in the CGM.
On the other hand, HM12 and FG09 have up to $\rm\sim$30 per cent of the total cold gas mass in the CGM for 
the Milky Way galaxy g8.26e11. Therefore, it is clear that LPF is effective in heating the halo gas and/or partially 
preventing it from cooling. For all three galaxies, the weaker UVB of FG09 results in larger cold gas masses than HM12 after $z\sim2$.  

In the same central panels of Figure~\ref{mass_tracks}, the dashed curves give the growth of the total 
HI mass inside $r_{\rm vir}$. These curves show that the LPF model (dashed red) leads to a slower growth of the HI mass in 
comparison to the corresponding UVB only model (dashed blue), for all three simulated galaxies. We plan to test in the future
this prediction regarding the HI mass against observations of the evolution of the cosmic HI density 
\citep[e.g.][]{Rhee:2018}, once we have a statistical sample of galaxies run with the LPF model.

Given the suppression in stellar mass induced by LPF, it is also interesting to look if there are any differences visible in the star 
formation rate (SFR) histories between the various radiation field models, which are shown in the bottom panel of Figure~\ref{mass_tracks}. 
The binning used to create these tracks is a constant 100~Myr. As already expected, there are very minor differences between HM12, FG09 and FG09+LPF in the SFR histories of the dwarf g1.08e11, but there are more notable ones for the Milky Way analogue g8.26e11 and the more massive galaxy g2.79e12. 
The galaxy g8.26e11 shows a slight SFR decrease in the LPF case as compared with the FG09 one, at almost all $z$s 
(solid red vs shaded blue). However, if we look at the SFR  histories in the very inner galaxy 
region $r<2\, \rm kpc$ (dashed curves), the LPF has a peak SFR of $\sim7\, \rm M_{\rm\odot}yr^{\rm -1}$, while FG09 reaches a peak of  $\sim10\, \rm M_{\rm\odot}yr^{\rm -1}$. For g2.79e12, which has its last important merger at $z\simeq1$, the LPF is particularly effective in reducing the post-merger SFR when compared with the UVB only case.    

To try discriminate whether the reduction in the cold gas mass within the galaxy region
is due to an in-situ heating by the LPF or to a slow 
down of  the accretion from the halo region, we computed the evolution of the accretion rate through a spherical shell of mean radius 
$0.2r_{\rm vir}(z)$ and thickness $0.02r_{\rm vir}(z)$. We chose this boundary definition to ensure that the cold 
gas disc (typically within $0.15r_{\rm vir}$)
is all inside this bounding sphere ($\lesssim0.20r_{\rm vir}(z)$) at any redshift. Next, we computed the radial velocities of all gas particles in this shell at all $z$s, and selected those that can cross the shell within 100~Myr such that we could directly compare the accretion rates with the SFR histories. 
The gas particles with negative radial velocities are considered to be inflowing, while those with positive velocities are outflowing. 
The evolution of the inflow and outflow rates, $\dot M_{\rm inflow}$ and $\dot M_{\rm outflow}$, for all three galaxies with all 
three heating models are shown as faint colored curves in the top panels of Figures~\ref{fig_accretion} and \ref{fig_outflow}, respectively. The faint colored curves in the bottom panels of the same two figures show 
the evolution of the median inflow and outflow gas temperatures, $T_{\rm inflow}$ and $T_{\rm outflow}$. 
To highlight the differences between the models, we also averaged the rates and temperatures on 500~Myr
time scales (solid colors).

Similar to the SFR histories, at a first look, the inflow rates shown in Figure~\ref{fig_accretion} are not much influenced by the 
choice of the heating/cooling model. At a closer look, however, the LPF model in the cases of the dwarf and the Milky Way leads to  
slightly smaller $\dot M_{\rm inflow}$ over the universe's lifespan than the FG09 one. The horizontal dotted lines in all panels 
give the median inflow rates over the complete galaxy histories, and represent a way of quantifying the small differences mentioned
before. All these values are listed in Table~\ref{table_z0properties}. For the dwarf, the median $\dot M_{\rm inflow}$ is the same 
for the LPF and the HM12 models, reinforcing the previous result from the stellar mass growth that a stronger background mimicks a 
weaker background plus the local photoionization field in the dwarf regime. For the most massive galaxy, $\dot M_{\rm inflow}$ does 
not distinguish among the three heating/cooling models. Closer to the current epoch, however, 
$\dot M_{\rm inflow}$ of the UVB+LPF model is visibly smaller than the rate in the UVB only case, for both the dwarf and 
the Milky Way-like galaxy. The numbers in color in each panel give the average $\dot M_{\rm inflow}$ over the last 4~Gyr. 
The more massive spiral suffered a major merger close to $z\sim1$, and shows an inverted trend, with the average late epoch 
inflow rate larger for the UVB+LPF than for the UVB only. Opposed to these very mild trends in inflow rate, the median inflow temperature histories in the bottom panels of Figure~\ref{fig_accretion} show large variations from model to model. For all three galaxies, the model with the weaker UVB, FG09, results in the smallest inflow temperatures at most redshifts, 
while the inclusion of the local photoionization field leads to the largest $T_{\rm inflow}$. 
This difference between the median $T_{\rm inflow}$ over the Universe's lifespan 
in the FG09 and LPF cases increases with the stellar mass, from $\rm\sim0.1dex$ for the dwarf to almost an order of magnitude 
for the most massive galaxy. Focusing only on the late times, the differences between LPF and FG09 in terms of 
average $T_{\rm inflow}$ are even larger, as given by the colored numbers in each panel.
Figure~\ref{fig_accretion} suggests that while the different heating/cooling models do not influence much the rate at which gas is 
accreted onto the galaxies, they do influence the temperature distributions of the inflow. 

Figure~\ref{fig_outflow} shows the equivalent of Figure~\ref{fig_accretion}, but for the outflowing gas. 
Similar to the inflow rates, the outflow rates $\dot M_{\rm outflow}$, which are driven by supernova feedback, 
are even less impacted by the choice of a particular heating/cooling model. Regarding the outflow temperature, 
the Milky Way galaxy shows the largest variation of almost $1dex$ between the low median $T_{\rm outflow}$ 
in the FG09 case, and the high LPF value. As opposed to the inflow temperature trends, the differences in the median 
$T_{\rm outflow}$ between FG09 and LPF do not increase with galaxy mass. Instead the most massive galaxy seems to 
be insensitive to the choice of heating/cooling model both in terms of outflow rates and outflow temperatures.
The temperatures of both inflowing and outflowing gas are larger for the stronger UVB of HM12 than for the FG09 one, 
at a fixed dark matter halo mass.

\begin{figure*}
\begin{center}
 \textbf{\underline{g1.08e11}}\\
\vspace*{0.2cm}
 
 \includegraphics[width=0.16\textwidth]{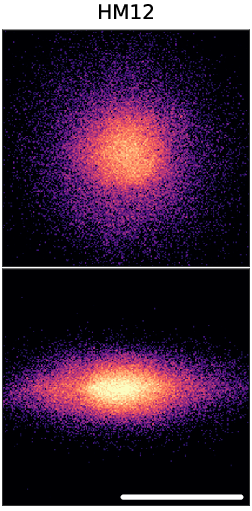}
 \includegraphics[width=0.16\textwidth]{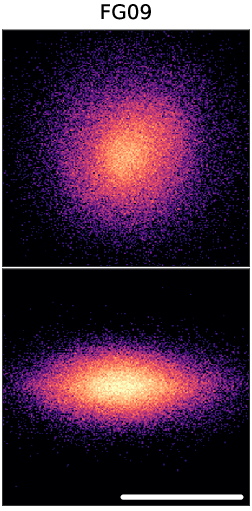}
 \includegraphics[width=0.16\textwidth]{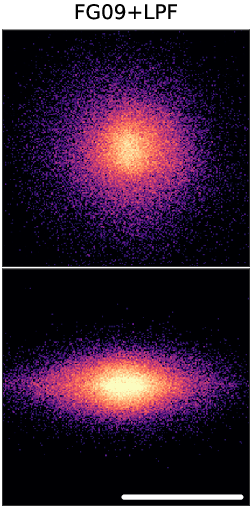}
 \includegraphics[width=0.16\textwidth]{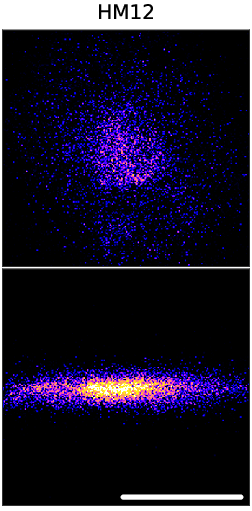}
 \includegraphics[width=0.16\textwidth]{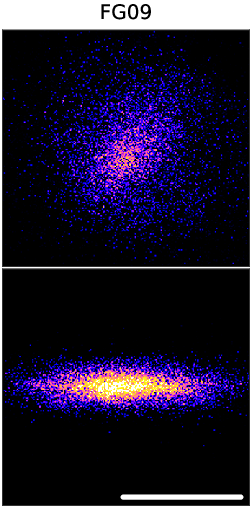}
 \includegraphics[width=0.16\textwidth]{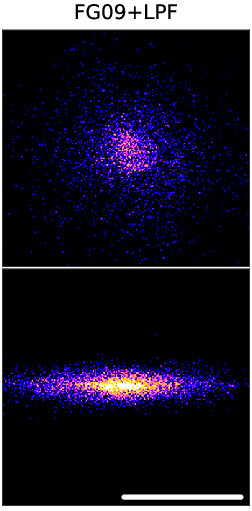}\\
 \includegraphics[width=0.49\textwidth]{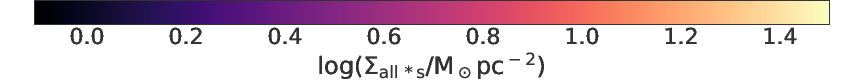}
 \includegraphics[width=0.49\textwidth]{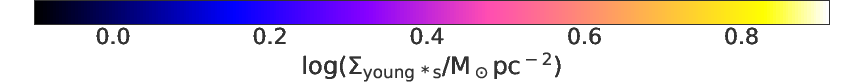}\\
 \vspace*{0.2cm}

 \textbf{\underline{g8.26e11}}\\
\vspace*{0.2cm}
 
 \includegraphics[width=0.16\textwidth]{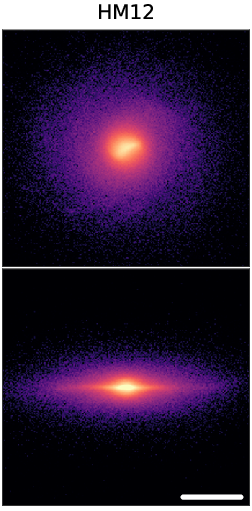}
 \includegraphics[width=0.16\textwidth]{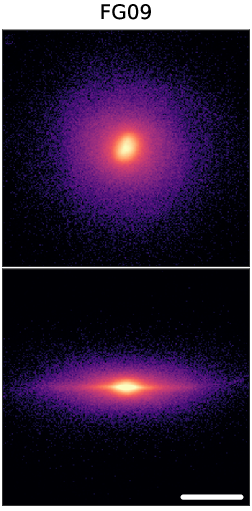}
 \includegraphics[width=0.16\textwidth]{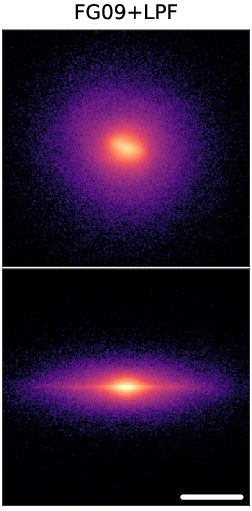}
 \includegraphics[width=0.16\textwidth]{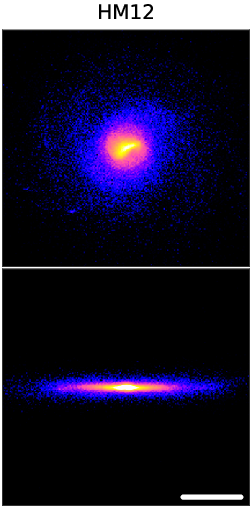}
 \includegraphics[width=0.16\textwidth]{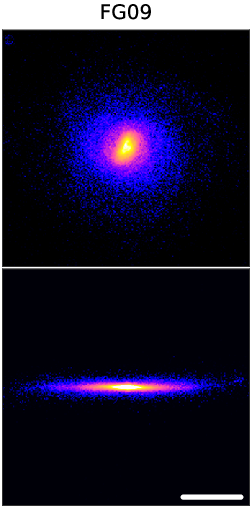}
 \includegraphics[width=0.16\textwidth]{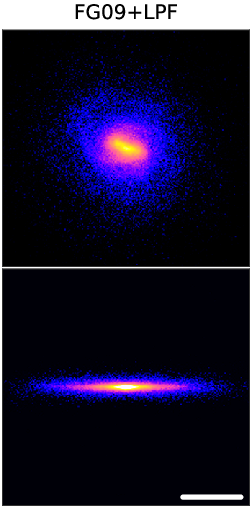}\\
 \includegraphics[width=0.49\textwidth]{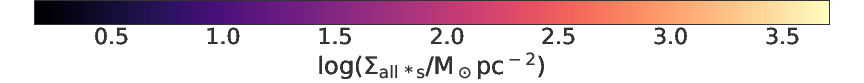}
 \includegraphics[width=0.49\textwidth]{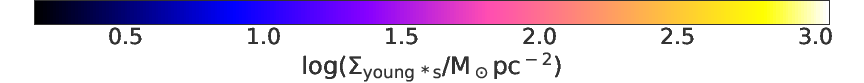}\\
\vspace*{0.2cm}

 \textbf{\underline{g2.79e12}}\\
\vspace*{0.2cm}
 
 \includegraphics[width=0.16\textwidth]{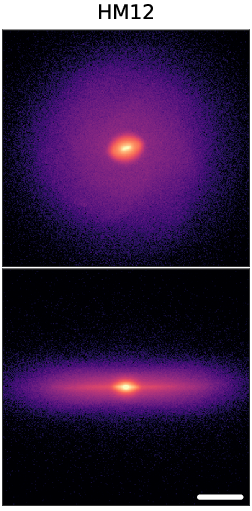}
 \includegraphics[width=0.16\textwidth]{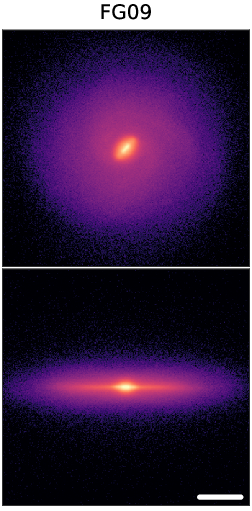}
 \includegraphics[width=0.16\textwidth]{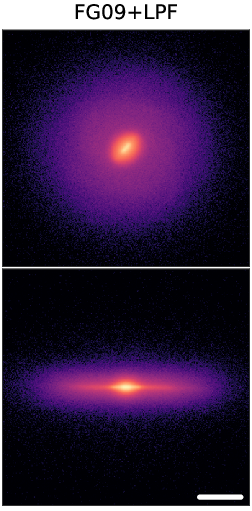}
 \includegraphics[width=0.16\textwidth]{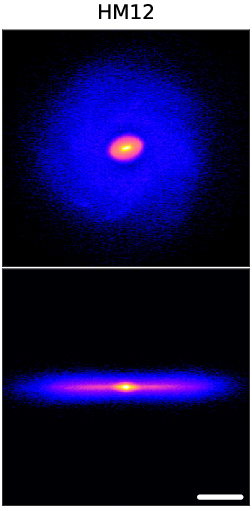}
 \includegraphics[width=0.16\textwidth]{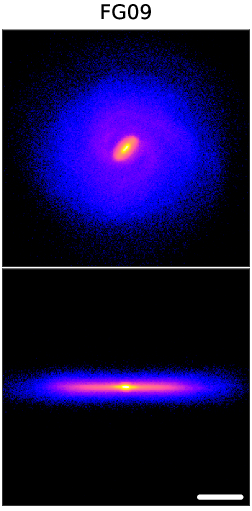}
 \includegraphics[width=0.16\textwidth]{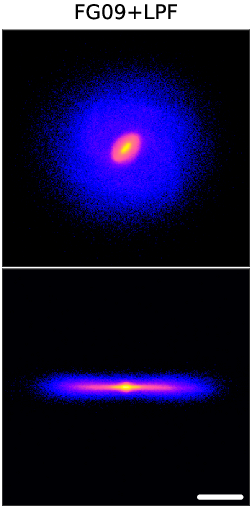}\\
 \includegraphics[width=0.49\textwidth]{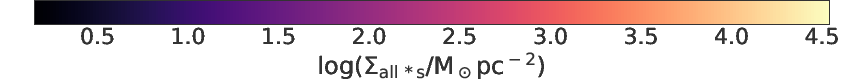}
 \includegraphics[width=0.49\textwidth]{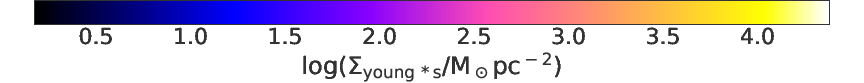}\\
 
 \caption{The stellar mass surface density maps in face-on and edge-on perspective. The left panels show all the stars, while the right ones give 
 only the maps corresponding to stars younger than 6 Gyr. The horizontal white bars represent the physical scale of 10 kpc.}
\label{fig_images_stars}
\end{center}
\end{figure*}

\subsection{Properties of galaxies at $z=0$}
\label{z0}

\subsubsection{Stellar population properties}
\label{z0_star}

The LPF is a preemptive feedback, and therefore is not expected to change significantly a galaxy's appearance. For a first test of 
the changes LPF can make in the $z=0$ galaxy's properties, Figure~\ref{fig_images_stars} gives the surface mass density maps for the stars.
This figure shows both the face-on and the edge-on surface mass density maps for the complete stellar populations 
(left panels) and for the stars younger than 6~Gyr (right panels). We choose this age cut because it roughly coincides with 
the epoch where the LPF starts producing an effect on the galaxy evolution (see Section~\ref{formation}).
For each galaxy and each component (either all stars or only young ones), the color bar is fixed for the three radiation field runs, 
such that the changes in mass surface density can be easier appreciated. To quantify the effects of the UVB and LPF on the stellar 
surface mass density maps, we use parameters of: S{\'e}rsic profiles \citep[effective surface mass density $\Sigma_{\rm e}$, 
effective radius $R_{\rm e}$ and S{\'e}rsic index $n$,][]{Sersic:1963}, exponential profiles (central surface mass density $\Sigma_{\rm 0}$
and scalelength $R_{\rm d}$), or both, fitted to the azimuthally averaged face-on maps.

\begin{figure*}
\begin{center}
 \includegraphics[width=0.31\textwidth]{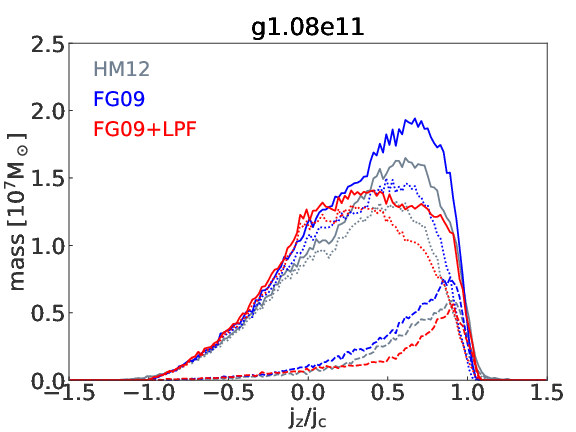}
 \includegraphics[width=0.31\textwidth]{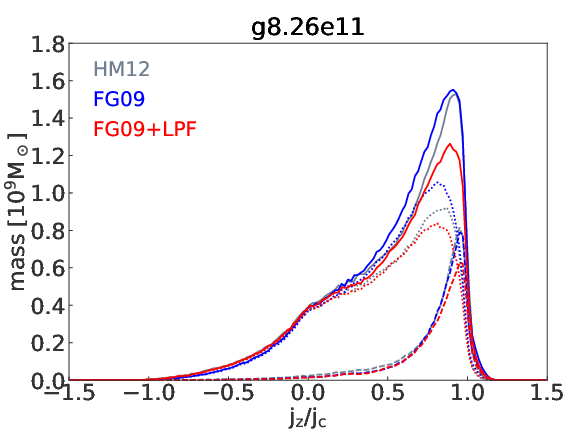}
 \includegraphics[width=0.31\textwidth]{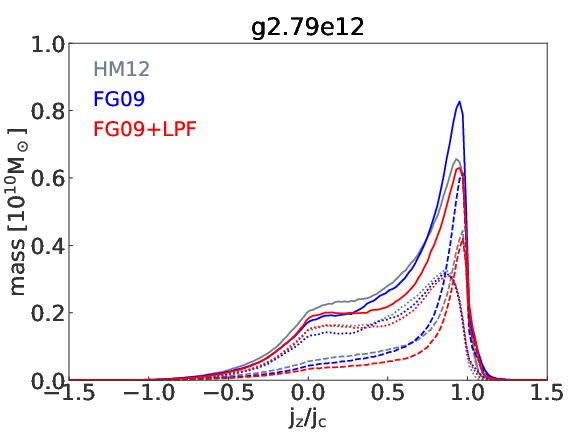}\\  
 \caption{Mass distributions of stellar particle circularities $j_{\rm z}/j_{\rm c}$ for the $z=0$ galaxies g1.08e11 (left), g8.26e11 (center) 
 and g2.79e12 (right), in the HM12 (grey), FG09 (blue) and FG09+LPF (red) runs. The dashed and dotted curves give the distributions for the stars 
 younger and older than 6~Gyr, respectively, while the solid curves represent the complete stellar populations.}
\label{fig_jzjc_z0}
\end{center}
\end{figure*}

For the MW galaxy g8.26e11, the complete stellar population is less centrally concentrated in the FG09+LPF model than in FG09, 
both in the edge-on and in the face-on projections. The face-on projection of g8.26e11 is well described by a single S{\'e}rsic profile.
The $\Sigma_{\rm e}$ of the FG09+LPF model is more than halfed with respect to that of FG09, $R_{\rm e}$ increases from $1.28\pm0.09$ to $1.67\pm0.07$ kpc, while $n$ decreases from $3.18\pm0.16$ to $2.82\pm0.10$. If we assume a fixed mass-to-light ratio $M/L$, in the 3.6$\rm\mu$m 
band for example, this 50\% difference in $\Sigma_{\rm e}$ translates into a surface brightness difference of 0.70~mag. 
The same is true when comparing the maps of the stars with ages $\rm<$6~Gyr, with a reduction in $\Sigma_{\rm e(<6Gyr)}$ of 32 per cent from 
FG09 to FG09+LPF, an increase in $R_{\rm e(<6Gyr)}$ from $1.35\pm0.08$ to $1.55\pm0.12$ kpc, and a decrease in $n_{\rm(<6Gyr)}$ from $2.80\pm0.12$ to $2.51\pm0.16$. When comparing the MW galaxy run with the two different UVB models, HM12 and FG09, there are no notable differences 
in the parameters of the S{\'e}rsic profiles of the young stellar populations, but more significant ones for the maps of all the stars, 
in the sense that HM12 results in a more centrally concentrated galaxy than FG09. All the parameters of these fits are given in Table~\ref{table_z0properties}. 

In all three models, the edge-on maps of the young stars of g8.26e11 seem to have a negligible bulge component, 
meaning that almost all stars younger than 6~Gyr belong to the $z=0$ disc. Given that this galaxy is well represented by 
a single S{\'e}rsic profile both for the complete stellar distribution as well as for the young stars, we can not use a 
\emph{photometric} bulge-to-total ratio ($B/T$) defined as the light/mass enclosed by the S{\'e}rsic profile divided by the 
total light/mass (S{\'e}rsic for the bulge + exponential for the disc).
However, we can quantify the $B/T$ of the stars via a \emph{dynamical} bulge-to-total ratio, and for this purpose we use the circularity parameter 
$j_{\rm z}/j_{\rm c}$ and the $B/T$ definition of \citet{Abadi:2003}. For each stellar particle, the circularity is computed as the azimuthal
(specific) angular momentum  $j_{\rm z}$ divided by the (specific) angular momentum of a particle with the same binding energy on a purely 
circular orbit $j_{\rm c}$. The distributions of $j_{\rm z}/j_{\rm c}$ for disc stars are strongly peaked at $j_{\rm z}/j_{\rm c}=1$, meaning 
that most of the stars are on circular orbits. The distributions for spheroid stars are typically symmetric and peaked around $j_{\rm z}/j_{\rm c}=0$,
if the spheroid has no net rotation. \citet{Abadi:2003} used these generalized characteristics of the $j_{\rm z}/j_{\rm c}$ distributions 
for discs and spheroids to define $B/T$ as:
\begin{equation}
 B/T = \frac{2\Sigma_{\rm k} m_{\rm k}(j_{\rm z}/j_{\rm c}<0)}{\Sigma_{\rm k} m_{\rm k}},
 \label{abadi_bd}
\end{equation}
where $m_{\rm k}$ is the mass of particle $k$. Figure~\ref{fig_jzjc_z0} gives the distributions of $j_{\rm z}/j_{\rm c}$ 
for the three galaxies simulated with the three radiation field models. 

The central panel of Figure~\ref{fig_jzjc_z0} shows the circularities distributions of the MW galaxy g8.26e11, splitted by stellar age:
the dashed curves represent the stars younger than 6~Gyr, while the dotted lines show stars older than this age cut. The solid lines are the 
circularity distributions for the complete stellar populations. In this figure we can see that indeed almost all young stars have 
$j_{\rm z}/j_{\rm c}>0$, with very small $B/T$ ratios of $0.06$ for HM12, $0.04$ for FG09, and $0.05$ for FG09+LPF 
(see Table~\ref{table_z0properties}). For the complete stellar populations of g8.26e11, the circularity distribution implies a decrease 
of the disc with respect to the bulge when the LPF is considered, $B/T=0.21$ for FG09 compared with $B/T=0.26$ for FG09+LPF. This seems 
a bit counterintuitive considering that the local photoionization field has the effect of flattening the surface mass density profile. 
To clarify this issue, one can look at how the spheroid stars are distributed. Considering that the dynamical $B/T$ is 
based on the assumption of a spherical symmetric spheroid, the radius enclosing half of the stellar mass on counter-rotating orbits 
($j_{\rm z}/j_{\rm c}\leqslant0$) should also be a good estimate for the half mass radius of the complete spheroid. Therefore, we 
computed the 2D half mass radius (in the plane of the disc) of the spheroid $R_{\rm 50}(j_{\rm z}/j_{\rm c}\leqslant0)$ 
for the three galaxies run with the three models. From the values of $R_{\rm 50}(j_{\rm z}/j_{\rm c}\leqslant0)$ in 
Table~\ref{table_z0properties}, the increase in $B/T$ from FG09 to FG09+LPF for all three galaxies can be explained by the increase 
in the size of the dynamical spheroid. Therefore, LPF increases the probability of stars at large galactocentric distances 
to be on counter-rotating orbits. In terms of $B/T$ ratios, the HM12 model is very close to FG+LPF for both the young and the old 
stellar populations.

The reduction in disc prominence is easily visible in the case of the dwarf galaxy g1.08e11. The surface mass density maps of the 
complete stellar populations in the top panels of Figure~\ref{fig_images_stars} show that the weaker UVB model of FG09 results in the less 
centrally saturated maps in both edge-on and face-on projections, when compared with both HM12 and FG09+LPF. This dwarf galaxy has a
face-on azimuthally average profile consistent with an exponential in all three UV models, and the biggest LPF effect is the reduction 
in scalelength $R_{\rm d}$ between FG09 and FG09+LPF, for both the complete (from $1.96\pm0.02$ to $1.80\pm0.01$ kpc) as well as the young 
(from $2.12\pm0.01$ to $1.87\pm0.02$ kpc) stellar populations. The HM12 model results in $R_{\rm d}$ in between FG09 and FG09+LPF, but 
closer to FG09+LPF. Therefore, from the profiles it is clear that LPF suppresses the disc growth (the normalization varies little among 
the three models). The same disc suppression effect is visible in the $j_{\rm z}/j_{\rm c}$ distributions of Figure~\ref{fig_jzjc_z0}, 
where $B/T$  increases from $0.44$ to $0.54$ between FG09 and FG09+LPF for the complete stellar populations, and from $0.15$ to $0.22$ 
for the young stars. When comparing the two UVB models, there are no significant differences between the $B/T$ ratio for any of the 
stellar populations considered. The more important effect is, however, the decrease in $R_{\rm d}$ when the UVB strength is increased:
from $1.96\pm0.02$ kpc for FG09 to $1.84\pm0.01$ kpc for HM12 for the complete stellar populations, and from $2.12\pm0.01$ to $1.90\pm0.02$ kpc
for the young stars. This is understandable considering the large differences in HI and He II photoionization rates between HM12 and FG09 
at $z\gtrsim1$, with $\Gamma_{\rm HI}\rm(HM12)$ being up to two times larger than $\Gamma_{\rm HI}\rm(FG09)$, and $\Gamma_{\rm HeII}\rm(HM12)$ 
up to almost one order of magnitude larger \citep[see Figure 4 in][]{Khaire:2019}.

\begin{table*}
\centering
\begin{tabular}{c|ccc|ccc|ccc}
\hline
Sim & & g1.08e11 & & & g8.26e11 & & & g2.79e12 & \\
\hline
Model & HM12 & FG09 & FG09+LPF & HM12 & FG09 & FG09+LPF & HM12 & FG09 & FG09+LPF\\
\hline
log($M_{\rm dark}$/M$_{\rm\odot}$) & 11.05 & 11.05 & 11.05 & 11.95 & 11.96 & 11.95 & 12.49 & 12.49 & 12.49\\
$r_{\rm vir}$ [kpc] & 105 & 104 & 105 & 213 & 214 & 213 & 322 & 321 & 322\\ 
log($M_{\rm *}$/M$_{\rm\odot}$) & 8.92 & 8.98 & 8.93 & 10.67 & 10.69 & 10.63 & 11.29 & 11.29 & 11.23\\
log($M_{\rm gas}$/M$_{\rm\odot}$) & 9.78 & 9.73 & 9.83 & 10.87 & 10.88 & 10.88 & 11.31 & 11.27 & 11.40\\
log($M_{\rm HI}$/M$_{\rm\odot}$) & 8.86 & 8.98 & 8.57 & 10.19 & 10.27 & 10.05 & 10.28 & 10.34 & 10.04\\
$R_{\rm HI}$ [kpc] & 5.84 & 6.24 & 7.02 & 26.72 & 33.57 & 23.60 & 24.02 & 23.49 & 18.35\\
$\dot M_{\rm inflow}$ [M$_{\rm\odot}$yr$^{\rm -1}$] & 0.52 & 0.59 & 0.52 & 2.30 & 2.72 & 2.26 & 7.47 & 6.61 & 6.92\\
$T_{\rm inflow}$ [K] & 1.6$\rm\times$10$^{\rm 4}$ & 1.4$\rm\times$10$^{\rm 4}$ & 1.8$\rm\times$10$^{\rm 4}$ & 1.5$\rm\times$10$^{\rm 4}$ & 1.3$\rm\times$10$^{\rm 4}$ & 2.7$\rm\times$10$^{\rm 4}$ & 2.2$\rm\times$10$^{\rm 4}$ & 1.9$\rm\times$10$^{\rm 4}$ & 1.7$\rm\times$10$^{\rm 5}$\\
$\dot M_{\rm outflow}$ [M$_{\rm\odot}$yr$^{\rm -1}$] & 0.32 & 0.34 & 0.26 & 1.11 & 1.15 & 1.17 & 3.61 & 3.78 & 4.23\\
$T_{\rm outflow}$ [K] & 1.7$\rm\times$10$^{\rm 4}$ & 1.4$\rm\times$10$^{\rm 4}$ & 1.8$\rm\times$10$^{\rm 4}$ & 2.5$\rm\times$10$^{\rm 4}$ & 1.6$\rm\times$10$^{\rm 4}$ & 1.1$\rm\times$10$^{\rm 5}$ & 3.5$\rm\times$10$^{\rm 6}$ & 3.0$\rm\times$10$^{\rm 6}$ & 3.1$\rm\times$10$^{\rm 6}$\\
log($\Sigma_{\rm e}$/M$_{\rm\odot}$pc$^{\rm -2}$) & - & - & - & 3.09$\rm\pm$0.05 & 3.21$\rm\pm$0.06 & 2.93$\rm\pm$0.04 & 1.66$\rm\pm$0.03 & 1.98$\rm\pm$0.04 & 1.85$\rm\pm$0.06\\
$R_{\rm e}$ [kpc] & - & - & - & 1.45$\rm\pm$0.08 & 1.28$\rm\pm$0.09 & 1.67$\rm\pm$0.07 & 10.60$\rm\pm$0.32 & 8.45$\rm\pm$0.30 & 8.08$\rm\pm$0.47\\
$n$ & - & - & - & 2.95$\rm\pm$0.12 & 3.18$\rm\pm$0.16 & 2.82$\rm\pm$0.10 & 0.87$\rm\pm$0.04 & 1.06$\rm\pm$0.05 & 1.21$\rm\pm$0.09\\
log($\Sigma_{\rm 0}$/M$_{\rm\odot}$pc$^{\rm -2}$) & 1.59$\rm\pm$0.01 & 1.56$\rm\pm$0.01 & 1.62$\rm\pm$0.01 & - & - & - & 4.68$\rm\pm$0.03 & 4.66$\rm\pm$0.03 & 4.43$\rm\pm$0.03\\
$R_{\rm d}$ [kpc] & 1.84$\rm\pm$0.01 & 1.96$\rm\pm$0.02 & 1.80$\rm\pm$0.01 & - & - & - & 0.67$\rm\pm$0.02 & 0.62$\rm\pm$0.02 & 0.80$\rm\pm$0.03\\
log($\Sigma_{\rm e(<6Gyr)}$/M$_{\rm\odot}$pc$^{\rm -2}$) & - & - & - & 2.52$\rm\pm$0.08 & 2.47$\rm\pm$0.05 & 2.30$\rm\pm$0.07 & 1.28$\rm\pm$0.04 & 1.70$\rm\pm$0.02 & 1.72$\rm\pm$0.11\\
$R_{\rm e(<6Gyr)}$ [kpc] & - & - & - & 1.33$\rm\pm$0.11 & 1.35$\rm\pm$0.08 & 1.55$\rm\pm$0.12 & 10.33$\rm\pm$0.35 & 8.98$\rm\pm$0.15 & 6.22$\rm\pm$0.53\\
$n_{\rm(<6Gyr)}$ & - & - & - & 2.88$\rm\pm$0.18 & 2.80$\rm\pm$0.12 & 2.51$\rm\pm$0.16 & 0.58$\rm\pm$0.03 & 0.62$\rm\pm$0.01 & 0.92$\rm\pm$0.07\\
log($\Sigma_{\rm 0(<6Gyr)}$/M$_{\rm\odot}$pc$^{\rm -2}$) & 0.85$\rm\pm$0.01 & 0.85$\rm\pm$0.01 & 0.80$\rm\pm$0.01 & - & - & - & 4.41$\rm\pm$0.05 & 4.39$\rm\pm$0.03 & 4.17$\rm\pm$0.09\\
$R_{\rm d(<6Gyr)}$ [kpc] & 1.90$\rm\pm$0.02 & 2.12$\rm\pm$0.01 & 1.87$\rm\pm$0.02 & - & - & - & 0.59$\rm\pm$0.02 & 0.53$\rm\pm$0.02 & 0.61$\rm\pm$0.06\\
$r_{\rm *50\%}$ [kpc] & 4.33 & 4.65 & 4.21 & 2.41 & 1.93 & 2.72 & 1.94 & 2.41 & 2.42\\
$conc_{\rm *}$=$\rm r_{\rm *90\%}/\rm r_{\rm *50\%}$ & 2.20 & 2.05 & 2.19 & 4.20 & 5.00 & 3.94 & 8.62 & 6.70 & 6.00\\
$V_{\rm c}^{\rm max}$ [km~s$^{\rm -1}$] & 92 & 93 & 94 & 250 & 268 & 231 & 498 & 470 & 420\\
$r(V_{\rm c}^{\rm max})$ [kpc] & 14.73 & 15.12 & 15.16 & 2.89 & 2.09 & 3.94 & 1.45 & 1.69 & 2.42\\
$B/T$ & 0.45 & 0.44 & 0.54 & 0.25 & 0.21 & 0.26 & 0.26 & 0.22 & 0.25\\
$B/T\rm(<6Gyr)$ & 0.13 & 0.15 & 0.22 & 0.06 & 0.04 & 0.05 & 0.22 & 0.14 & 0.16\\
$B/T\rm(\geqslant6Gyr)$ & 0.53 & 0.51 & 0.59 & 0.31 & 0.25 & 0.32 & 0.29 & 0.28 & 0.30\\
$R_{\rm 50}(j_{\rm z}/j_{\rm c}\leqslant0)$ [kpc] & 2.64 & 2.90 & 2.55 & 1.21 & 0.90 & 1.42 & 0.89 & 0.95 & 1.23\\
$f_{\rm cold}$ & 0.51 & 0.65 & 0.28 & 0.56 & 0.59 & 0.30 & 0.21 & 0.24 & 0.09\\
$f_{\rm hot}$ & 0 & 0 & 0 & 0.33 & 0.30 & 0.51 & 0.72 & 0.68 & 0.88\\
$h_{\rm HI,50}$ [kpc] & 0.16 & 0.26 & 0.12 & 0.24 & 0.30 & 0.21 & 0.32 & 0.39 & 0.21\\
$h_{\rm HI,90}$ [kpc] & 0.54 & 0.90 & 0.26 & 0.89 & 1.96 & 0.79 & 1.22 & 1.38 & 0.61\\
$\sigma_{\rm z,*bulge}$ [km~s$^{\rm -1}$] & 28$\rm\pm$1 & 29$\rm\pm$1 & 29$\rm\pm$1 & 110$\rm\pm$10 & 114$\rm\pm$14 & 100$\rm\pm$7 & 222$\rm\pm$28 & 212$\rm\pm$30 & 201$\rm\pm$22\\
$\sigma_{\rm z,*disc}$ [km~s$^{\rm -1}$] & 14$\rm\pm$5 & 16$\rm\pm$5 & 14$\rm\pm$5 & 72$\rm\pm$9 & 74$\rm\pm$10 & 70$\rm\pm$8 & 105$\rm\pm$26 & 91$\rm\pm$20 & 100$\rm\pm$19\\
\hline
\end{tabular}
\centering\caption{Global properties for galaxies g1.08e11, g8.26e11 and g2.79e12 simulated with the three models for the radiation field, HM12, FG09 
and FG09+LPF. All properties except for the inflow/outflow rates and temperatures refer to the galaxies at $z=0$. 
$M_{\rm dark}$, $M_{\rm *}$, $M_{\rm gas}$ and $M_{\rm HI}$ are the dark matter, stellar, gas and HI masses inside the virial radius 
$r_{\rm vir}$, while $R_{\rm HI}$ is the radius at which the face-on HI surface mass density drops below 
1~M$_{\rm\odot}$pc$^{\rm -2}$. 
$\dot M_{\rm inflow}$/$\dot M_{\rm outflow}$ are the median inflow/outflow rates through a shell of radius $0.2r_{\rm vir}$ 
and thickness $0.02r_{\rm vir}$ through the universe's lifetime, while $T_{\rm inflow}$/$T_{\rm outflow}$ are the respective median 
inflow/ouflow temperatures. $\Sigma_{\rm e}$, $R_{\rm e}$ and $n$ are the effective surface mass densities, effective radii and 
S{\'e}rsic indices from the fits to the azimuthally averaged face-on stellar surface mass density maps, while $\Sigma_{\rm e(<6Gyr)}$,
$R_{\rm e(<6Gyr)}$, $n_{\rm(<6Gyr)}$ are the parameters of the corresponding fits to the maps of stars younger than 6~Gyr. 
The parameters of the exponential fits are the central surface mass density $\Sigma_{\rm 0}$ and scalelength $R_{\rm d}$ for the 
complete stellar populations, and $\Sigma_{\rm 0(<6Gyr)}$ and $R_{\rm d(<6Gyr)}$ for the young stars. The 3D stellar half mass radii 
and concentrations are $r_{\rm *50\%}$ and $conc_{\rm *}$=$\rm r_{\rm *90\%}/\rm r_{\rm *50\%}$. $V_{\rm c}^{\rm max}$ is the maximum
circular velocity (considering all mass), and $r(V_{\rm c}^{\rm max})$ is the radius at which the maximum circular velocity occurs. 
The \emph{dynamical} bulge-to-total ratios for the complete, young and old stellar populations are $B/T$, $B/T\rm(<6Gyr)$ and 
$B/T\rm(\geqslant6Gyr)$, while $R_{\rm 50}(j_{\rm z}/j_{\rm c}\leqslant0)$ is the 2D radius in the disc plane enclosing half of the stellar mass on counter-rotating orbits ($j_{\rm z}/j_{\rm c}\leqslant0$). 
The fractions of cold ($<$15000~K) and hot ($>$80000~K for g8.26e11, and $>$160000~K for g2.79e12) gas 
inside $r_{\rm vir}$ are $f_{\rm cold}$ and $f_{\rm hot}$. The heights above the symmetry plane enclosing 
50 and 90 per cent of the HI disc mass are $h_{\rm HI,50}$ and $h_{\rm HI,90}$. 
The stellar bulge and disc vertical velocity dispersion are $\sigma_{\rm z,*bulge}$ 
and $\sigma_{\rm z,*disc}$, the first being computed within $r_{\rm *50\%}$ and the latter within $r_{\rm *50\%}<R<r_{\rm *90\%}$.}
\label{table_z0properties}
\end{table*}

The more massive galaxy g2.79e12 shows much more the effect of LPF in the stellar surface mass density maps of Figure~\ref{fig_images_stars}. 
This galaxy has a conspicuous bar in all three runs, whose strength seems to decrease from the FG09 model to FG09+LPF. However, 
the most significant difference can be seen in the extent of the stellar disc, especially when comparing the younger stellar maps 
of FG09 and FG09+LPF. As opposed to the MW galaxy g8.26e11, the younger stellar populations of g2.79e12 are not confined to the disc only, 
having a clear contribution in the central bar region as well. Due to the presence of the bar, this galaxy is not well fitted by a single 
S{\'e}rsic profile, requiring two components. Therefore, we use a combined S{\'e}rsic+exponential to fit the azimuthally averaged face-on maps. 
Interestingly, for this galaxy the central part is well described by the exponential and the disc by the S{\'e}rsic, contrary to what is typically
assumed in observational studies. The parameters of the fits for both the complete and the young stellar populations can be found in 
Table~\ref{table_z0properties}. 

As suggested by Figure~\ref{fig_images_stars}, the LPF has the effect of damping the central surface mass density 
peak of g2.79e12, which translates into a decrease in $\Sigma_{\rm 0}$ and an increase in $R_{\rm d}$ between FG09 and FG09+LPF, and 
a suppression of the stellar disc, exemplified by a decrease in both $\Sigma_{\rm e}$ and $R_{\rm e}$ between FG09 and FG09+LPF. 
The disc profiles for the complete stellar distributions are actually close to exponentials ($n\sim1$), but we did not fix the index 
because the young stellar populations require $n<1$. When comparing the two UVB models, there are no significant differences between the 
inner region described by ($\Sigma_{\rm 0}$,$R_{\rm d}$) in neither the case of the complete stellar population, nor for the young stars.
However, there are significant differences between the discs. For the complete stellar populations, the most notable difference is between 
the normalizations of the disc, with $\Sigma_{\rm e}{\rm(FG09)}=2\times\Sigma_{\rm e}{\rm(HM12)}$. For the young stellar populations the 
difference in the disc normalization is even bigger: $\Sigma_{\rm e}{\rm(FG09)}=2.6\times\Sigma_{\rm e}{\rm(HM12)}$. 
Similar to g8.26e11, the \emph{dynamical} bulge-to-total ratio increases from FG09 to FG09+LPF, though only by $\rm\sim$2 per cent for any of 
the stellar populations considered. The massive galaxy g2.79e12 is $\rm\sim$75 per cent disc dominated in the FG09+LPF, and only slightly more 
($\rm\sim$78 per cent) in the FG09 case. This galaxy shows a big difference in $B/T$ of old stars between the run with the HM12 background and 
the one with the FG09, the former resulting in an almost twice as large $B/T\rm(<6Gyr)$: $0.22$ versus $0.14$. 

It is interesting also to compare the $B/T$ ratios defined as in photometry with the \emph{dynamical} values. For this purpose, we can integrate 
the mass under the exponential central part $M_{\rm bulge} = 2\pi\Sigma_{\rm 0}R_{\rm d}^{\rm 2}$ and divide it by the corresponding total 
stellar mass. In this manner we obtain the ratios of $0.69$, $0.57$ and $0.63$ for the HM12, FG09 and FG09+LPF models, respectively. If we 
would use this definition of bulge-to-total mass ratio, we would conclude that g2.79e12 is clearly bulge dominated, as opposed to being 
disc dominated as suggested by the \emph{dynamical} $B/T$ ratios, which are $0.26$, $0.22$ and $0.25$, respectively.  

To quantify the thickness of the galaxies' stellar components, we compute both the bulge and disc vertical velocity dispersions 
$\sigma_{\rm z,*bulge}$ and $\sigma_{\rm z,*disc}$ (see Table~\ref{table_z0properties}). The bulge vertical velocity dispersion 
is calculated within the 3D stellar half mass radius $r_{\rm *50\%}$, while the disc's is computed within $r_{\rm *50\%}<R<r_{\rm *90\%}$.
For the dwarf galaxy g1.08e11, all three models for heating/cooling lead to the same $\sigma_{\rm z,*bulge}$ and $\sigma_{\rm z,*disc}$
within the errors. For the L$^{\rm *}$ galaxies g8.26e11 and g2.79e12, LPF results in a decrease in $\sigma_{\rm z,*bulge}$ and an
increase in $\sigma_{\rm z,*disc}$, when comparing FG09 with FG09+LPF. The two different UVB models seem to have an effect only on g2.79e12, 
with the weaker UVB of FG09 resulting in smaller vertical velocity dispersions for both bulge and disc than HM12.

\begin{figure}
\begin{center}
 \textbf{\underline{g1.08e11}}\\
\vspace*{0.2cm}
 
 \includegraphics[width=0.15\textwidth]{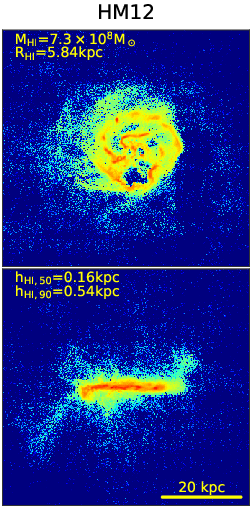}
 \includegraphics[width=0.15\textwidth]{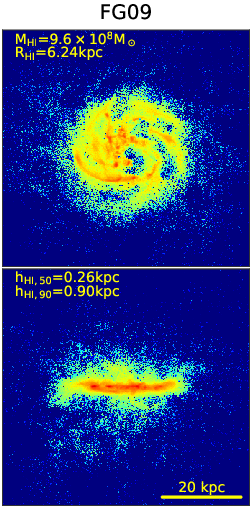}
 \includegraphics[width=0.15\textwidth]{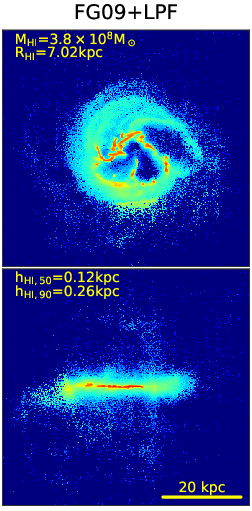}\\
  \includegraphics[width=0.45\textwidth]{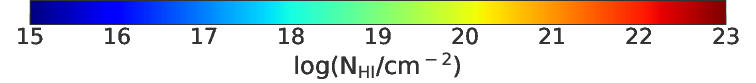}\\
\vspace*{0.2cm}

 \textbf{\underline{g8.26e11}}\\
\vspace*{0.2cm}
 
 \includegraphics[width=0.15\textwidth]{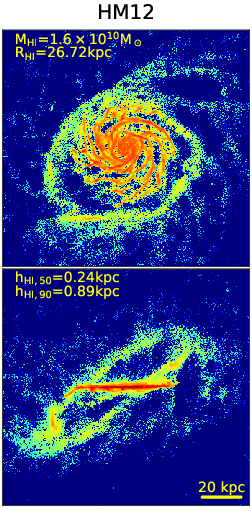}
 \includegraphics[width=0.15\textwidth]{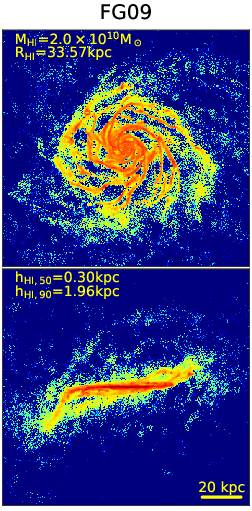}
 \includegraphics[width=0.15\textwidth]{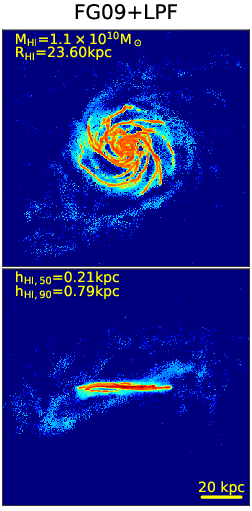}\\
  \includegraphics[width=0.45\textwidth]{fig_7h_new.png}\\
\vspace*{0.2cm}

 \textbf{\underline{g2.79e12}}\\
\vspace*{0.2cm}
 
 \includegraphics[width=0.15\textwidth]{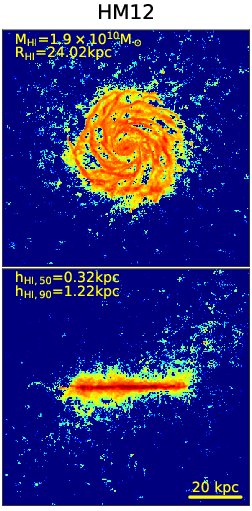}
 \includegraphics[width=0.15\textwidth]{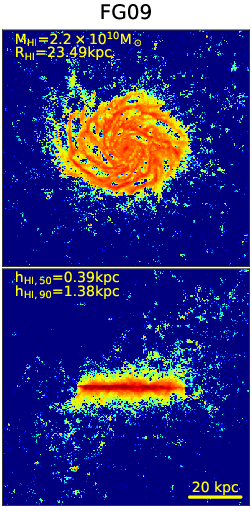}
 \includegraphics[width=0.15\textwidth]{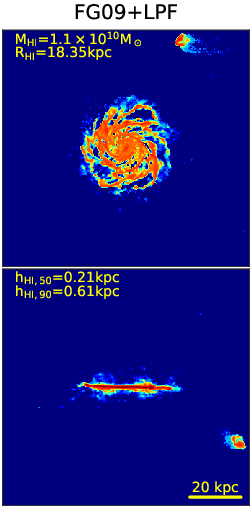}\\
 \includegraphics[width=0.45\textwidth]{fig_7h_new.png}\\
 
 \caption{The HI column density maps in face-on and edge-on perspective. The pixel scales are given by the 
 corresponding gas force softenings of Table~\ref{table_runs}. The text on the face-on panels gives the total HI mass in the region 
 shown, as well as the HI radius, while the text on the edge-on maps shows the HI half mass and 90\% heights.}
\label{fig_images_gas}
\end{center}
\end{figure}

\subsubsection{Gas properties}
\label{z0_gas}

The large differences in the evolution of the cold gas masses among the three heating/cooling models seen in Figure~\ref{mass_tracks}
already suggests that there should be much more variance in the gas morphology at $z=0$ than in the stars. Therefore, in 
Figure~\ref{fig_images_gas} we show the face-on and edge-on column density maps for the HI gas in the galaxy.

All three galaxies show a reduction in the thickness of the HI discs from the FG09 model to FG09+LPF, 
as quantified by the HI half mass height $h_{\rm HI,50}$ given in each edge-on panel. The two more massive galaxies, g8.26e11 and g2.79e12, also show a significant decrease in the HI disc extent, as quantified by the HI radius $R_{\rm HI}$ of the face-on 
projections. This radius is defined as the radius where the mass profile drops below 1~M$_{\rm\odot}$pc$^{\rm -2}$.
We compute the 50 and 90\% HI heights using only the pixels with $R<R_{\rm HI}$.  

The LPF does not impact significantly the extent of the HI disc in the dwarf galaxy g1.08e11, 
but is does reduce its vertical extent by half, from $0.26$~kpc for FG09 to $0.12$~kpc for FG09+LPF. 
The large bubbles visible in the face-on maps of g1.08e11 are caused by SNe II explosions. The stochastic nature of the SNe II leads 
to the different bubble morphology's, which should not be used as a criteria to distinguish among the three heating/cooling models.    
The reduction between FG09 and LPF in HI disc thickness are large for all three galaxies, ranging from 
30\% for the $h_{\rm HI,50}$ of g8.26e11 to 46\% for g2.79e12, and from 56\% for the $h_{\rm HI,50}$ of g2.79e12 to 71\% for g1.08e11. 
Also, the HI total masses drop by $\rm\sim$50\% between FG09 and LPF for all three galaxies, as was already shown in 
Figure~\ref{mass_tracks}. However, it is important to note that the cold gas disc of 
g8.26e11 in all three models is slightly warped (see the edge-on projections in Figure~\ref{fig_images_gas}). 

When comparing the cold gas morphologies in the HM12 and FG09 models in Figure~\ref{fig_images_gas},  the 
stronger UVB leads to a thinner HI gas disc. The dwarf and the Milky Way-like galaxy 
are also less extended, as quantified by $R_{\rm HI}$, in the HM12 model than in the FG09 one. On the other hand, the more 
massive object g2.79e12 actually shows a slight increase of $0.5$ kpc in $R_{\rm HI}$ between FG09 and HM12.

\begin{figure*}
\begin{center}
 \includegraphics[width=0.31\textwidth]{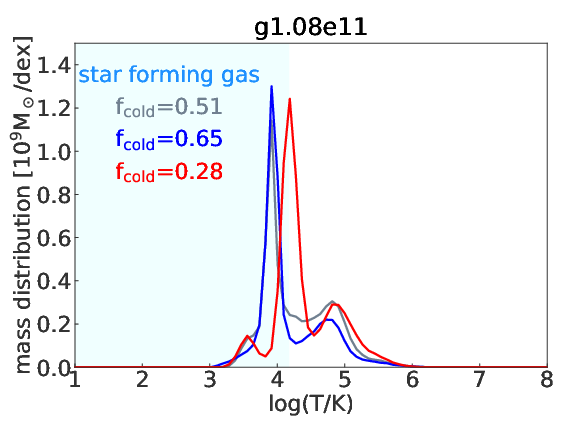}
 \includegraphics[width=0.31\textwidth]{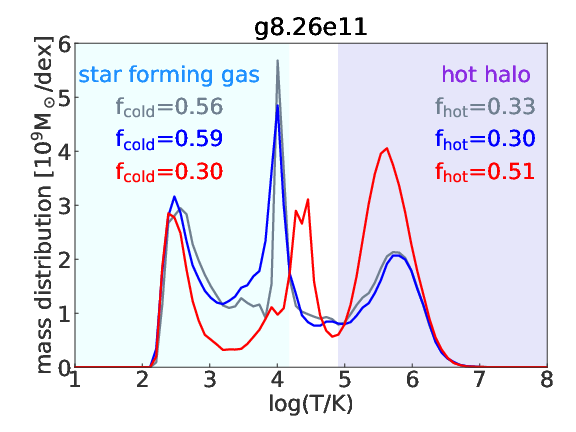}
 \includegraphics[width=0.31\textwidth]{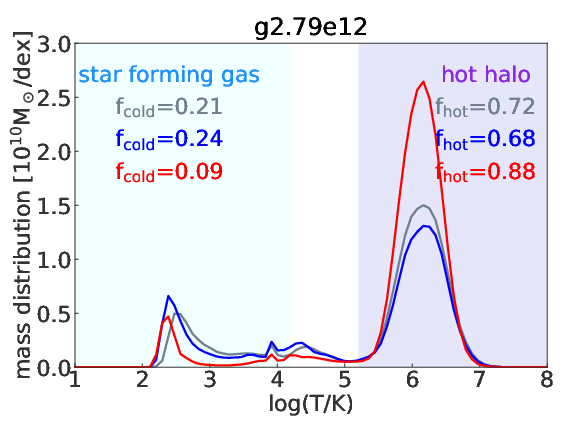}\\   
 \caption{The distributions of gas temperatures in the HM12 (grey), FG09 (blue) and FG09+LPF (red) runs 
 for the $z=0$ galaxies g1.08e11 (left), g8.26e11 (center) and g2.79e12 (right). The light cyan shaded areas mark the cold 
 (T$\rm<$15000~K) gas phase, which is the reservoir for star formation. The light purple shaded areas define the hot halo gas phase, 
 which corresponds to T$\rm>$80000~K for g8.26e11, and to T$\rm>$160000~K for g2.79e12. The dwarf galaxy g1.08e11 has no significant 
 hot halo phase. The colored numbers to the left/right of each panel give the corresponding gas mass fractions in the cold/hot gas phases.}
\label{fig_tdistz0}
\end{center}
\end{figure*}

\begin{figure*}
\begin{center}
 \includegraphics[width=0.31\textwidth]{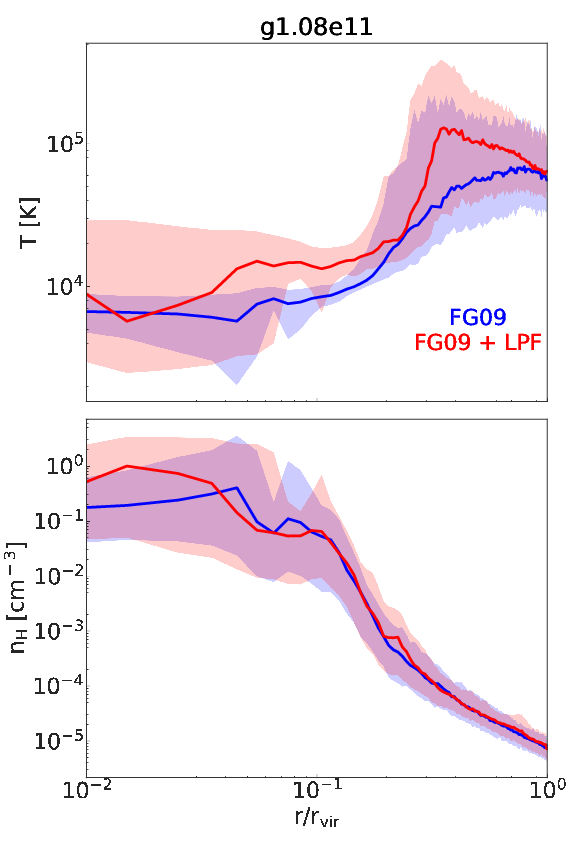}
 \includegraphics[width=0.31\textwidth]{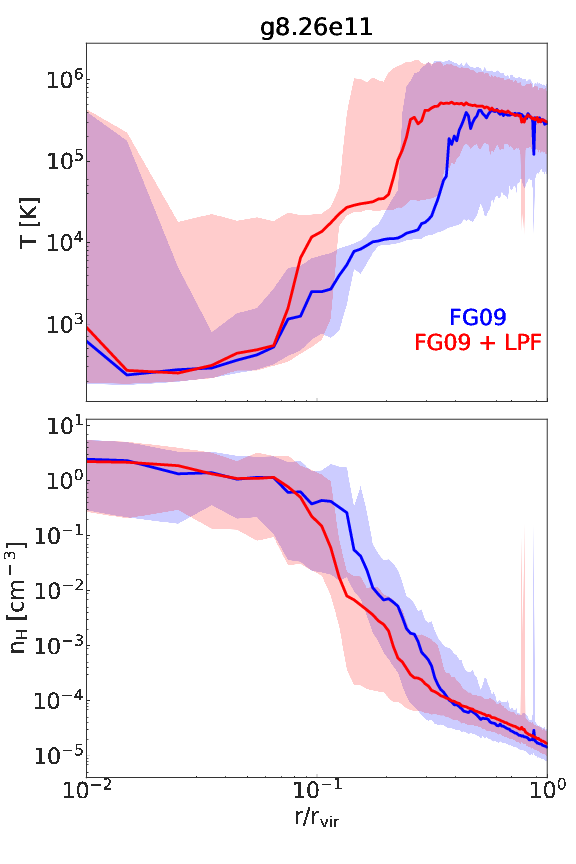}
 \includegraphics[width=0.31\textwidth]{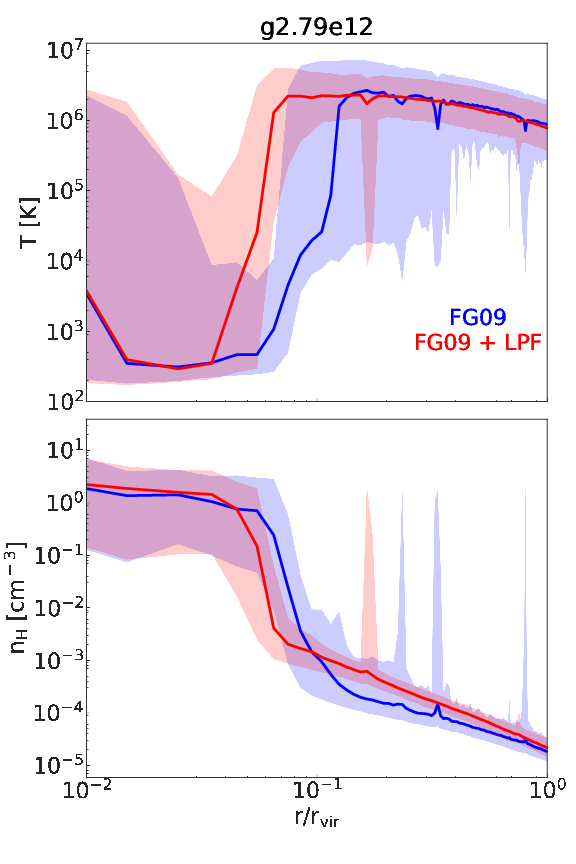}\\  
 \caption{The temperature (top) and density (bottom) structure of the gas as a function of radius in the FG09 (blue) and 
 FG09+LPF (red) runs for the $z=0$ galaxies g1.08e11 (left), g8.26e11 (center) and g2.79e12 (right). The shaded 
 regions mark the 10th to the 90th per cent quantiles, while the solid curves give the 50 per cent quantiles. 
 In the case of the two more massive galaxies, 
 note how the hot halo boundary moves towards smaller radii when the LPF is included. 
 The HM12 model is very similar to FG09, and therefore is not plotted.}
\label{fig_tnprof}
\end{center}
\end{figure*}

\begin{figure*}
\begin{center}
 \includegraphics[width=0.31\textwidth]{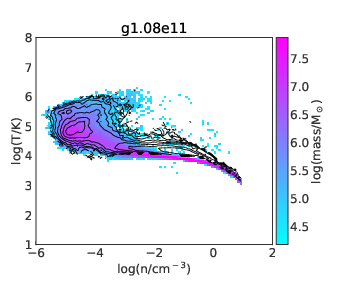}
 \includegraphics[width=0.31\textwidth]{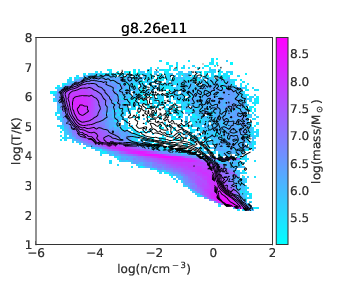}
 \includegraphics[width=0.31\textwidth]{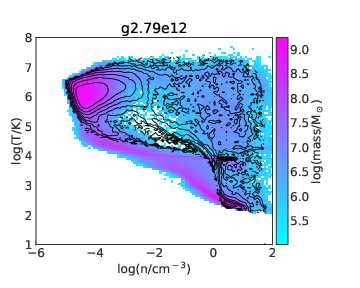}\\  
 \caption{The phase diagrams of the FG09+LPF (black contours) models compared with the FG09 (2D color map) ones
 for the $z=0$ galaxies g1.08e11 (left), g8.26e11 (center) and g2.79e12 (right).}
\label{fig_phasediag_z0}
\end{center}
\end{figure*}

\begin{figure}
\begin{center}
 \includegraphics[width=0.45\textwidth]{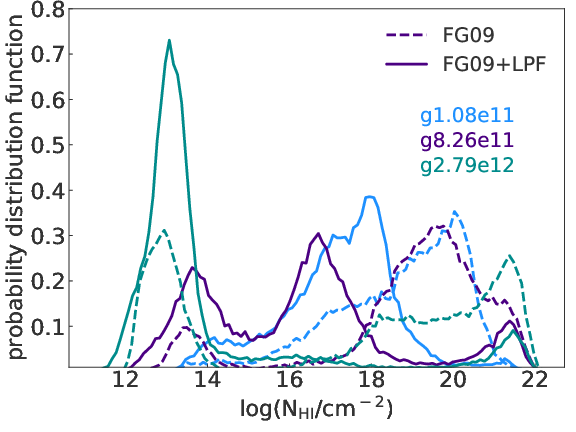}\\  
 \caption{Comparison between the distributions of HI column densities of the FG09 model (dashed curves) and 
 those of FG09+LPF (solid curves) at $z=0$. The distributions have been constructed from the face-on images shown in 
 Figure~\ref{fig_images_gas}, and therefore cover fully the ISM and partially the CGM of the galaxies. 
 Both inclinations and resolution have an effect on these distributions.}
\label{fig_NHI_pdfs}
\end{center}
\end{figure}

An important effect of the LPF is visible in the general structure of the gas in the virial sphere, and in particular in the 
temperature distribution functions, as shown in Figure~\ref{fig_tdistz0}. The light cyan and purple shaded areas in Figure~\ref{fig_tdistz0} 
mark the cold and hot gas phases. While we consider the cold gas phase to be fixed by the limiting temperature for star formation 
(T$\rm<$15000~K), the hot phase will generally vary with the dark matter halo mass. As it can be seen from Figure~\ref{fig_tdistz0}, 
the dwarf galaxy shows no peak at high temperatures, while the two most massive galaxies do. The limiting temperatures for the hot 
gas phase of the  galaxies g8.26e11 and g2.79e12 are 80000~K and 160000~K, respectively. This limiting 
temperature for each
galaxy has been chosen based on the dip between the peak of the hot halo and the rest of the gas temperature distribution. 
While there are only small differences between the HM12 and FG09 runs in any of the three galaxies, the LPF show a significant 
decrease of the cold gas phases with respect to FG09 for all three galaxies, and an important increase of the hot gas phase for 
the two most massive galaxies. This increase/decrease in $f_{\rm hot}$/$f_{\rm cold}$ between FG09 and LPF is 
due to the extra photoheating produced by the local radiation field.
For the cases of the dwarf g1.08e11 and the Milky Way analogue g8.26e11, the cold gas fractions 
$f_{\rm cold}$ are halfed between the FG09 and FG09+LPF models, while the decrease is even higher for the massive galaxy g2.79e12, 
$f_{\rm cold}=0.24$ as compared to $f_{\rm cold}=0.09$. The two more massive galaxies compensate the decrease in $f_{\rm cold}$ by a 
significant increase in the fraction of hot halo gas $f_{\rm hot}$ between FG09 and FG09+LPF, from $0.30$ to $0.51$ for g8.26e11, 
and from $0.68$ to $0.88$ for g2.79e12. The total gas masses (within the virial radius) for each simulation are given in 
Table~\ref{table_z0properties}, together with the cold and hot phase fractions, $f_{\rm cold}$ and $f_{\rm hot}$, respectively. 
The two massive galaxies have an important hot halo contribution in the temperature range 10$^{\rm 5.5}\rm<$T$\rm<$10$^{\rm 6.5}$,
which give part of the Bremsstrahlung emission taken into account by the LPF (see Section~\ref{section_lpf}). 
The most massive galaxy also has some gas in the second temperature bin used for the Bremsstrahlung emission. 
None of the three galaxies has any hot halo gas above 10$^{\rm 7.5}$K. 

\begin{figure*}
\begin{center}
 \includegraphics[width=0.31\textwidth]{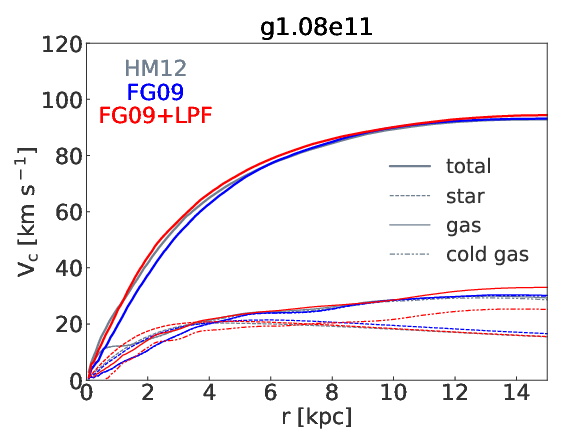}
 \includegraphics[width=0.31\textwidth]{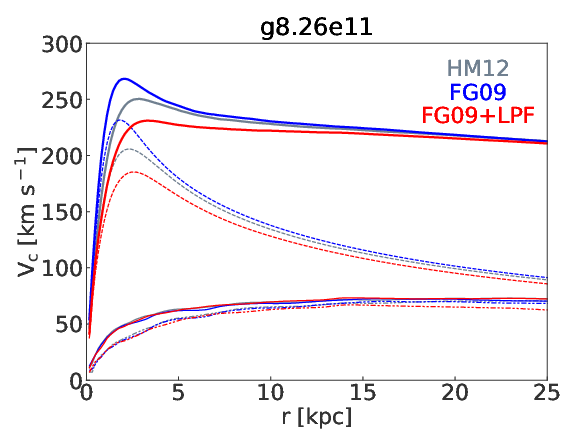}
 \includegraphics[width=0.31\textwidth]{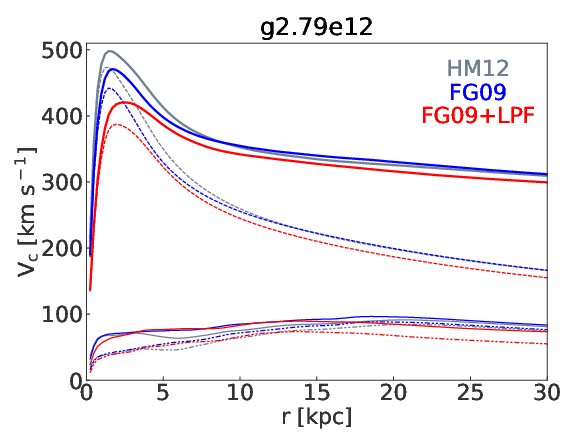}\\ 
 \caption{The circular velocity profiles in the HM12 (grey), FG09 (blue) and FG09+LPF (red) runs 
 for the $z=0$ galaxies g1.08e11 (left), g8.26e11 (center) and g2.79e12 (right). The thick curves give the total circular velocity, while the 
 dashed, solid and dotted-dashed curves show the contributions of the stars, gas and cold gas, respectively.}
\label{fig_vcprof_z0}
\end{center}
\end{figure*}

\begin{figure*}
\begin{center}
 \includegraphics[width=0.31\textwidth]{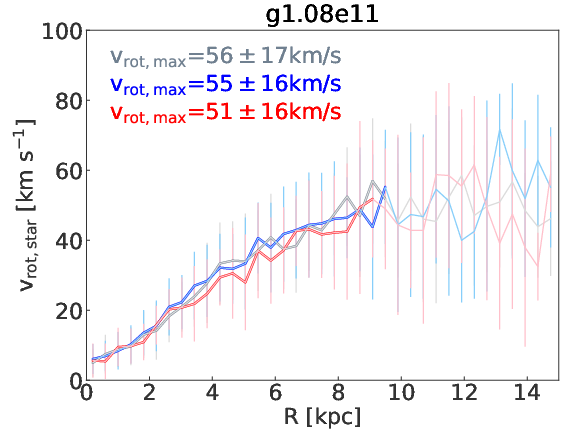}
 \includegraphics[width=0.31\textwidth]{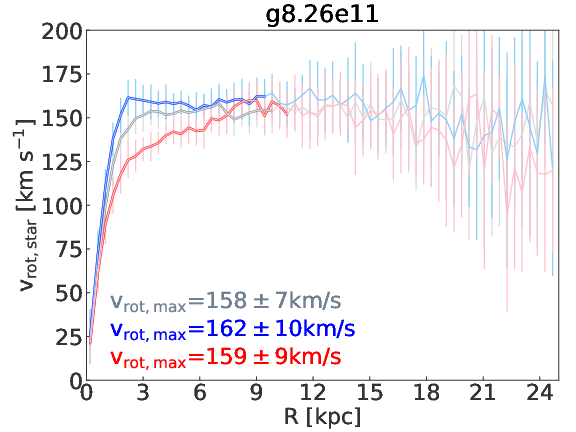}
 \includegraphics[width=0.31\textwidth]{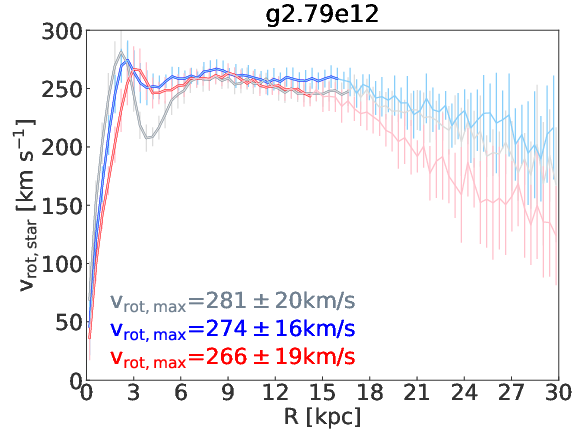}\\

 \includegraphics[width=0.31\textwidth]{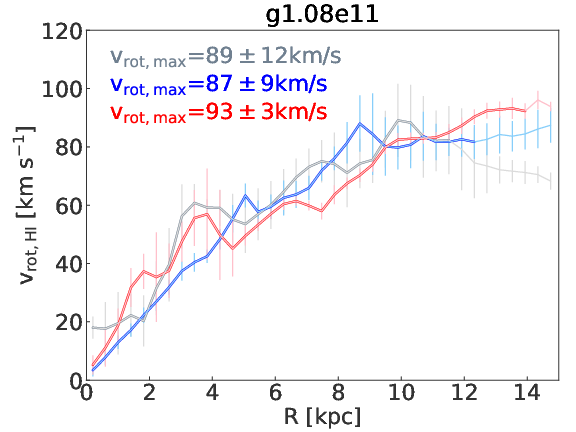}
 \includegraphics[width=0.31\textwidth]{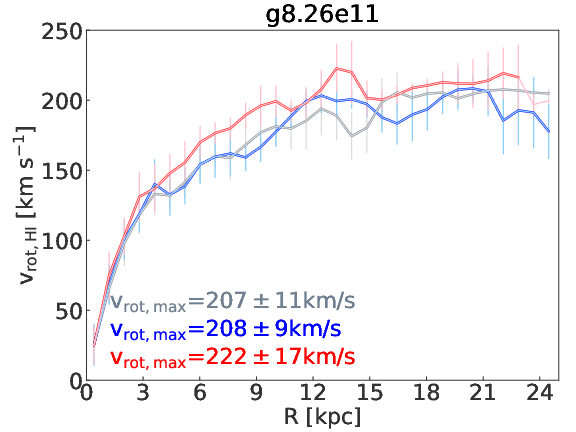}
 \includegraphics[width=0.31\textwidth]{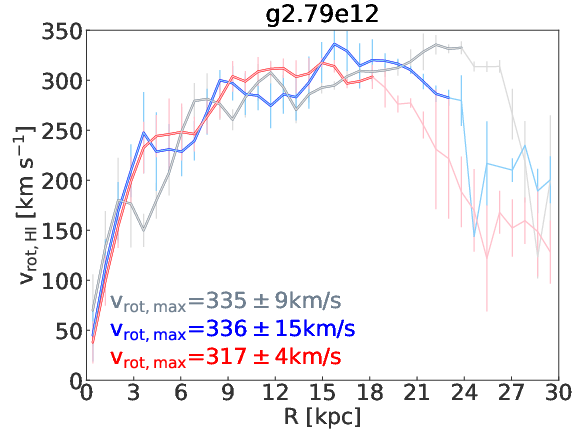}\\

 \caption{The line-of-sight edge-on rotational velocity profiles along the semi-major axis of the three $z=0$ galaxies run with the HM12 (grey), FG09 (blue) and LPF (red) models. The top panels show the stellar velocity profiles, while the bottom ones give the 
 HI gas.  The transparent sections of the curves mark the regions outside the corresponding 90 per cent enclosed mass radius for the stars and $R_{\rm HI}$ for the gas, respectively. For the dwarf, the limiting radius is 
 2$R_{\rm HI}$ in the bottom panel.  The transparent error bars are the mass weighted standard deviation of the line-of-sight velocities.}
\label{fig_appendix2}
\end{center}
\end{figure*}

To complement the global temperature distributions, Figure~\ref{fig_tnprof} gives the radial structure in both temperature  
and density for the three galaxies run with the three models for the radiation field. The median temperature and density 
as a function of radius are shown by the solid curves, while the corresponding regions between the 10th and 90th 
per cent quantiles are given as shaded areas.
As already hinted from Figure~\ref{fig_tdistz0}, the differences in radial temperature and density profiles between HM12 and FG09 are small, and therefore we do not show HM12. 
When comparing the FG09 runs with their corresponding LPF ones, it is easy to observe that the boundary between galaxy and hot 
halo gas is significantly more inward in the LPF case. Another important difference between FG09 and LPF is 
particularly visible in the case of the most massive galaxy, where the 10 per cent quantile for the temperature profile 
(lower limit of the shaded area) at intermediate radii 
(0.1$\rm\lesssim$r/r$_{\rm vir}\rm\lesssim$0.4) is $\rm\sim10^{\rm 6}$~K for LPF compared with $\rm\sim10^{\rm 4}$~K for FG09. 
This means that in the g2.79e12 LPF case, there is very little gas in the halo (less than 10 per cent) at temperatures 
$\rm<10^{\rm 6}$~K. If this lack of $\rm\sim10^{\rm 4}$~K gas in the haloes of massive galaxies at $z=0$ proves to be statistically 
robust (more simulations are needed to test this hypothesis), upper limits or detections of Lyman-$\rm\alpha$ in absorption 
along QSOs sightlines at various galactocentric radii of low $z$ galaxies \citep[see review by][]{Tumlinson:2017}
could be used as a discriminant between different feedback models like the ones discussed in this work.  
It is also important to note than none of the three galaxies shows big differences in terms of the gas density and temperature 
structure at $z=0$ between the two UVB models, HM12 and FG09. At higher $z$s, it is likely that the virial gas structure will 
show an imprint of the metagalactic background strength. In a future study we will present the radial distribution of 
the column density of various ion species for a larger sample of galaxies and compare the simulations with and without 
LPF with observational data.

A global view on the differences induced by the effects of the local photoionization/photoheating field on the gas is given by 
the phase space diagrams. These diagrams are shown in Figure~\ref{fig_phasediag_z0}, where the FG09 models for the three galaxies
given by the 2D color maps are compared with the corresponding FG09+LPF ones given by the black contours. 
The phase diagram of the dwarf galaxy g1.08e11 is not much impacted by LPF, the only effect appearing as a slight 
increase in the equilibrium temperature for densities $n\rm\gtrsim$10$^{\rm -3}$cm$^{\rm -3}$. The effect of LPF is much more significant 
for the two more massive galaxies, where the minimum temperature is pushed above $\sim10^{\rm 4.5}$~K (FG09+LPF)
from the typical value of $\rm\sim10^{\rm 4}$~K (FG09) in the density range $\rm10^{\rm -4}\lesssim n \lesssim10^{\rm 0}cm^{\rm -3}$.
At densities higher than 1~cm$^{\rm -3}$, the FG09 and FG09+LPF models converge because the densest gas is shielded against the radiation 
field (see Section~\ref{fig_phasediag_z0}). The equilibrium temperature $T_{\rm eq}\gtrsim10^{\rm 4.5}$~K in the FG09+LPF runs 
is close to the temperature where cooling is most efficient for both primordial and solar metallicity gas \citep[e.g.][]{Wiersma:2009}. 
This fact together with the density based shielding from the radiation field explains why there is not a more dramatic effect of 
LPF at L$^{\rm *}$ galaxy mass scales. 

A more robust test of the LPF predictions in terms of HI is to compare it with observational measurements 
of the cosmic distribution function of HI column densities \citep[e.g.][]{Zwaan:2005}, and with the $z$-evolution of cosmic HI density 
\citep[e.g.][]{Rhee:2018}. Such comparisons, however, require statistical samples of galaxies, and therefore will be tackled in 
a future work. Nevertheless, we can already look at how the distributions of HI column densities vary between the UVB only model 
and LPF, as shown in Figure~\ref{fig_NHI_pdfs}. For all three galaxies, LPF moves the peak(s) of the $N_{\rm HI}$ distributions to 
lower values, and the more massive the galaxy the bigger the effect. The differences between the UVB only and LPF
models are large such that observational inferred HI cosmic densities, which are typically column density limited 
\citep[e.g. $N_{\rm HI}\rm>$10$^{\rm 19}$cm$^{\rm -2}$ for the observations of][]{Zwaan:2005}, could 
discriminate between the two models.

\subsubsection{Velocity curves}

One way to compare the mass distributions of the various runs is by means of the circular velocity $V_{\rm c}(r)=\sqrt{\rm GM(<r)/r}$ 
profiles. Therefore, Figure~\ref{fig_vcprof_z0} shows the various contributions to the circular velocity profiles. 
The total $\rm V_{\rm c}$ (dark matter + gas + stars) in given by the thick solid curves, while the solid, 
dashed and dotted-dashed represent the gas, stars and cold gas contributions, respectively. As in the previous figures, 
the dwarf galaxy is mostly insensitive to the changes in the radiation field implementation, having a slowly rising circular 
velocity curve in all three heating/cooling models. Contrasting, the two more massive galaxies show a large drop in the peak 
circular velocity between the UVB only and the LPF cases, due mainly to an important reduction in the stellar mass in the inner 
regions. For the Milky Way and the most massive galaxies, the peak circular velocities drop by $\rm\sim$40~km~s$^{\rm -1}$ and 
$\rm\sim$50~km~s$^{\rm -1}$ between the FG09 and LPF, and the positions of the peaks move outwards by 1.8~kpc and 0.7~kpc, 
respectively (see Table~\ref{table_z0properties} for the exact values). 
In general, at all $z$s, the SFR is more radially spread, and lower for the LPF than for the UVB 
only models, especially in the very inner regions. This effect is caused by the lower levels of star forming gas, caused at their turn 
by the LPF induced photoheating. This fact explains why the peak in $V_{\rm c}$ at $z=0$ is moving outwards. The drop in maximum 
$V_{\rm c}$ is mainly driven by the global drop of $\rm\sim$20\% in final stellar mass, but also by the more spread SFR throughout the disc. All three galaxies show differences in the cold gas contribution to $V_{\rm c}(r)$ at large radii. 

In observations what is usually measured is not $V_{\rm c}$, but the line-of-sight velocities along the major axis 
of a galaxy. Therefore, in Figure~\ref{fig_appendix2} we give these line-of-sight velocities measured along the major 
axis of the galaxies in edge-on perspective, and we call them rotational velocities. The top panels of this figure show 
the stellar velocities, while the bottom ones the cold gas. The numbers in each panel give the maximum velocities of each model.

The first impression from Figure~\ref{fig_appendix2} is that the various heating/cooling models do not differ much 
in terms of $v_{\rm rot,HI}$. In terms of stellar rotational velocities, the dwarf galaxy is also insensitive to
the radiation field model, but the two more massive galaxies show significant differences at smaller radii, converging 
to the same value at large $R$. The Milky Way galaxy g8.26e11, e.g., passes from having a very sharp increase in 
$v_{\rm rot,star}$ followed by a plateau in the FG09 model to a more slowly rising rotational velocity curve for FG09+LPF. 
For this galaxy, the UVB model of HM12 results in a rotational velocity curve in between the ones of FG09 and FG09+LPF. 
At a closer look, there are differences also between $v_{\rm rot,HI}$ profiles of the FG09 and FG09+LPF for g8.26e11, 
with FG09+LPF resulting in higher rotational velocities at most radii. The more massive galaxy g2.79e12 shows an interesting 
effect in $v_{\rm rot,star}$, with the local radiation field dumping to a great extent the features associated with the presence 
of the bar at $R<$6~kpc. The HM12 model for this galaxy shows a significant dip in $v_{\rm rot,star}$ at the radius where the 
bar ends $R\sim$4~kpc \citep[see ][ for an analysis of the bar features in the higher resolution version of g2.79e12]{Buck:2018}.  

One interesting fact emerging from the comparison between Figure~\ref{fig_vcprof_z0} and Figure~\ref{fig_appendix2} is 
how different in both shape and normalization are $V_{\rm c}$, $v_{\rm rot,HI}$ and $v_{\rm rot,star}$. The shape of 
the stellar rotational velocity, for example, gives little indication on how peaked is the circular velocity profile. 
On the other hand, the HI rotational velocityis much more slowly increasing than $v_{\rm rot,star}$, which at its 
turn is increasing more slowly than $V_{\rm c}$, irrespective of the galaxy total mass. 

\section{Summary and discussion}
\label{conclusions}

We have implemented the optically thin local radiation field model of \citet{Kannan:2014,Kannan:2016} in the new version of the N-body 
SPH code {\sc gasoline2}, and use it to study how the local photoionization feedback (LPF) affects galaxy formation in cosmological simulations. 
As test cases we have used three NIHAO galaxies: one dwarf ($M_{\rm dark}\rm\simeq10^{\rm 11}M_{\rm\odot}$), one Milky Way analogue 
($M_{\rm dark}\rm\simeq10^{\rm 12}M_{\rm\odot}$) and one massive spiral ($M_{\rm dark}\rm\simeq10^{\rm 12.5}M_{\rm\odot}$). 
The model considers three types of local sources on top of a $z$-dependent homogeneous UVB: young massive stars and SNe remnants, 
post-AGB stars, and Bremsstrahlung from hot halo gas. The model of Kannan et al. has been constructed on top of the UVB of FG09, 
while the original NIHAO suite has been simulated with the HM12 UVB. Therefore, we also study the differences between the same 
three galaxies run with the two different UVBs. 

In terms of mass budget, the choice of UVB results in different  $M_{\rm *}$ only for the dwarf galaxy 
(decrease by 13 per cent between FG09 and HM12), while the final stellar masses of the more massive galaxies remain unchanged. 
The cold gas mass fractions do not change significantly between the two UVBs for the two more massive galaxies, but the 
larger ionizing photon flux in the HM12 model results in smaller HI masses by 17 and 13 per cent for g8.26e11 and g2.79e12, respectively.
As for $M_{\rm *}$, the choice of UVB has a larger impact on the dwarf galaxy, the cold gas fraction $f_{\rm cold}$ decreasing from 
0.65 for FG09 to 0.51 for HM12, while $M_{\rm HI}$ decreases by 24 per cent between the two models. For the two more massive galaxies, 
the biggest differences between FG09 and HM12 appear in the edge-on structure of the HI discs (Figure~\ref{fig_images_gas}), while in face-on projection the Milky Way galaxy has clearly a larger extent when simulated with the model which has less UV radiation capable of ionizing the gas, i.e. FG09.

From the comparison between the galaxies run with and without LPF, it results that LPF: 
\begin{itemize}
 \item increases the median temperature (averaged over the universe's lifespan) of the gas inflowing towards the galaxy from 0.11~dex 
 for the dwarf galaxy up to 0.95~dex for the massive spiral, without affecting significantly the gas inflow rate. 
 \item suppresses the star formation, such that at $z=0$ the total stellar mass is lower by $\rm\sim$20 per cent for all three galaxies.
 \item keeps a significant gas fraction above the temperature threshold for star formation, this fraction 
 increasing with time and with dark matter halo mass. The fraction of hot halo gas is increased by $\rm\sim$20 per cent for the two more 
 massive galaxies, while the fraction of cold gas is decreased by 50 per cent or more for all three objects.
 \item increases the \emph{dynamical} bulge-to-total ratio at $z=0$ by a few percent (10 per cent for the dwarf, 5 per cent for the Milky Way, 
 and 3 per cent for the massive spiral). 
 \item moves inward the border between the hot halo and the galaxy in the case of the two more massive objects. This translates into 
 significantly lower HI disc scalelengths and thicknesses, and less cold gas in the region of the halo.
 \item lowers the peak of the total circular velocity profile for the two more massive galaxies. In the case of the Milky Way-like galaxy,  $V_{\rm c}$ passes from peaked to flat. For the more massive spiral, LPF also lowers $V_{\rm c}^{\rm max}$ by 50~km~s$^{\rm -1}$, but not enough  to result in a flat circular velocity curve.
 \item decreases the gas fraction $f_{\rm gas}=M_{\rm HI}/(M_{\rm HI}+M_{\rm *})$ by 40 per cent for the dwarf and the 
 massive spiral, and by 24 per cent for the Milky Way.
 \item moves the distributions of $N_{\rm HI}$ towards lower column density values, and the effect is bigger the 
 more massive the galaxy.
 \item only partially alleviates the overcooling problem of the massive spiral galaxy; the central stellar mass density is decreased and the half mass radius is increased. It damps the strength of the bar of this galaxy to produce a significant effect in the 
 line-of-sight (edge-on projection) stellar velocity curve along the major axis. 
\end{itemize}

In the light of these differences, it is clear that LPF does affect L$^{\rm *}$ galaxies. However, the local radiation field does not 
provide enough feedback to solve the overcooling problem for galaxies more massive than the Milky Way, for which AGN feedback is thought 
to be the solution. 

For the Milky Way-like galaxy, which is a good Galaxy analogue in terms of solar neighborhood disc structure 
\citep{Obreja:2018}, we derive the UV (1500 \AA) interstellar radiation field from the LPF model at $R$=8~kpc, and find a value of the same order of magnitude as measurements for the Galaxy \citep{Witt:1973,Henry:1980,vanDishoeck:1994}.

The largest impact of this type of preemptive feedback is on the temperature structure of the gas in the virial sphere, 
both at the HI disc scales, as well as in the CGM.
We plan to refine this model both in terms of radiation sources taken into account and chemistry 
(inclusion of H$_{\rm 2}$ and of a more detailed metal line cooling), and test it against observations in the nearby universe. 

The detailed gas phase space distribution in galaxies and their haloes is the most promising approach to distinguish between various 
feedback models (stellar, radiation and AGN). In this respect, a larger sample of simulated galaxies could already provide the means to 
refine/distinguish feedback models, when tested against observations of nearby galaxies in terms of HI disc morphology and rotational 
pattern \citep[e.g.][]{Walter:2008}, HI disc power spectrum \citep{Grisdale:2017}, 
cosmological HI column density distribution \citep[e.g.][]{Zwaan:2005,Martin:2010,Braun:2012,Delhaize:2013},
equivalent widths of various ion species at various galactocentric distances 
\citep[e.g.][and references therein]{Tumlinson:2017}, and X-ray emission from the hot haloes of massive spiral galaxies \citep[e.g.][]{Bogdan:2013a,Bogdan:2013b}. 

As we saw from our simulations, the evolution of the amount of HI gas in the halo also differs strongly between the UVB only and the UVB+LPF models, and therefore a powerful observational constrain is provided by the evolution of the
HI cosmic density \citep[e.g.][]{Rao:2006,Prochaska:2009,Noterdaeme:2012,Zafar:2013,Rhee:2013,Rhee:2018}. For this test, however, we 
first have to build a statistical sample of galaxies. 

Last, but not least, any model of the simulated local radiation field can be tested 
against observations of the interstellar radiation field of the Galaxy \citep[e.g.][]{Popescu:2017}, and/or against cosmic UV emissivities \citep[e.g.][]{Sawicki:2006,Yoshida:2006,Bouwens:2007,Reddy:2008}.

\section{Acknowledgments}

We would like to thank Fabrizio Arrigoni Battaia for insightfull discussions related to this manuscript. 
This research was carried out on the High Performance Computing resources at New York University Abu Dhabi; 
on the \textsc{theo} cluster of the Max-Planck-Institut f\"{u}r Astronomie and on the \textsc{hydra} clusters at the Rechenzentrum in Garching. 
We greatly appreciate the contributions of these computing allocations.
All figures in this work have been made with {\tt matplotlib} \citep{Hunter:2007}, using the open-source 
Python libraries: {\tt pynbody} \citep{Pontzen:2013}, {\tt numpy} \citep{Walt:2011} and {\tt scipy} \citep{Jones:2001}. 
AO and BM are funded by the Deutsche Forschungsgemeinschaft (DFG, German Research Foundation) -- MO 2979/1-1.
TB acknowledges support from by the European Research Council under ERC- CoG grant CRAGSMAN-646955.
RK acknowledges support from NASA through Einstein Postdoctoral Fellowship grant number PF7-180163 awarded by the Chandra X-ray Center, 
which is operated by the Smithsonian Astrophysical Observatory for NASA under contract NAS8-03060.

\bibliographystyle{mnras}
\bibliography{lpf_rev}

\end{document}